%% file: main.tex
\documentclass[a4paper,11pt]{article}
\pdfoutput=1 

\usepackage{jcappub} 

\usepackage[T1]{fontenc} 
\usepackage[normalem]{ulem}
\usepackage{comment}
\usepackage{listings}

\def\be{\begin{equation}}
\def\ee{\end{equation}}
\def\bea{\begin{eqnarray}}
\def\eea{\end{eqnarray}}

\def\gev{\, {\rm GeV}}
\def\mev{\, {\rm MeV}}
\def\tev{\, {\rm TeV}}

\def\cm{\, {\rm cm}}
\def\km{\, {\rm km}}

\def\s{\, {\rm s}}

\title{Prospects for measuring dark matter microphysics with observations of dwarf spheroidal galaxies}

\author[a]{Eric J.~Baxter}
\affiliation[a]{\mbox{Institute for Astronomy, University of Hawai'i, 2680 Woodlawn Drive, Honolulu, HI 96822, USA}}
\emailAdd{ebax@hawaii.edu}

\author[b]{Jason Kumar}
\affiliation[b]{\mbox{Department of Physics \& Astronomy,
University of Hawai'i, Honolulu, HI 96822, USA}}
\emailAdd{jkumar@hawaii.edu}

\author[c]{Andrew B.~Pace}
\affiliation[c]{McWilliams Center for Cosmology, Carnegie Mellon University, 5000 Forbes Avenue, Pittsburgh, PA 15213, USA}
\emailAdd{apace@andrew.cmu.edu}

\author[b]{Jack Runburg}
\emailAdd{runburg@hawaii.edu}

\abstract{Dark matter annihilation in dwarf spheroidal (dSph) galaxies near the Milky Way has the potential to produce a detectable signature in gamma-rays.  The amplitude of this signal depends on the dark matter density in a dSph, the dark matter particle mass, the number of photons produced in an annihilation, and the possibly velocity-dependent dark matter annihilation cross section.  We argue that if the amplitude of the annihilation signal from multiple dSphs can be measured, it is possible to determine the velocity-dependence of the annihilation cross section.  However, we show that doing so will require improved constraints on the dSph density profiles,  including control of possible sources of systematic uncertainty.  Making reasonable assumptions about future improvements, we make forecasts for the ability of current and future experiments --- including Fermi, CTA and AMEGO --- to constrain the dark matter annihilation velocity dependence.}

\begin{document}
\maketitle
\flushbottom

\section{Introduction}

A key strategy for studying dark matter is the search 
for photons arising from dark matter annihilation in 
dwarf spheroidal galaxies (dSphs).  dSphs are promising 
search targets because they are thought to be dark 
matter-dominated astrophysical objects with 
relatively small astrophysical foregrounds.   
Searches for dark matter annihilation in dSphs have 
thus far yielded tight bounds, but no significant 
evidence of a signal (e.g. \cite{Fermi:2015, Archambault:2017}).  It is hoped that, as more dSphs are found, 
and as they are studied with instruments probing new 
energy ranges with larger exposures, evidence for dark 
matter annihilation may yet be forthcoming.  In this 
paper, we investigate a related question: if future observations with gamma-ray 
telescopes find evidence for 
dark matter annihilation in dSphs, can these observations also be used to determine the velocity-dependence of the microscopic dark matter annihilation process? 
 
The flux of photons arising from dark matter annihilation 
in any astrophysical object is proportional to the object's 
$J$-factor, which encodes all of the dependence of the 
photon flux on the astrophysical details of the target.  
The $J$-factors are typically determined by analyzing 
stellar velocity data, which can be used to infer the dSph mass distribution.  Recent work has demonstrated that 
these $J$-factors depend non-trivially on the 
velocity-dependence of the dark matter annihilation 
cross section~\cite{Robertson:2009bh,Belotsky:2014doa,Ferrer:2013cla,Boddy:2017vpe,Zhao:2017dln,Petac:2018gue,Boddy:2018ike,Lacroix:2018qqh,Boddy:2019wfg,Boddy:2020}.  

For unresolved observations of the photon flux from a single dSph,  information about the velocity-dependence encoded in the $J$-factor will be degenerate with the annihilation cross section, particle mass, and the number of photons produced per annihilation, which also impact the expected photon flux.  However, for a set of dSphs with different characteristic dark matter velocities, changing the velocity-dependence of the annihilation cross section will impact the $J$-factor of each dSph differently.  Consequently, gamma-ray observations of multiple dSphs can be used to break degeneracies between the annihilation velocity dependence and other quantities that impact the amplitude of the annihilation signal. 

Such an analysis will also be impacted by a variety of additional sources of uncertainty.  First, for any choice of 
velocity-dependence, the calculation of the $J$-factors 
from stellar data is plagued by parameter degeneracies, which can significantly degrade constraints \cite[e.g.][]{Bonnivard:2015, Pace:2018tin}.  Secondly, astrophysical 
foregrounds can complicate the determination of the 
photon flux arising from dark matter annihilation.  
But as more stars in a dSph are observed, and with 
greater precision, the uncertainties in the $J$-factors 
are expected to decrease.  By the same token, as 
more dSphs are found, and as observations are made with 
larger exposures, the statistical impact of the foregrounds will decrease.  Finally, there are several potential sources of systematic uncertainty that may impact $J$-factor
constraints, such as differences between the true dark matter profile and the assumed profile used in the stellar analysis \citep{Bonnivard:2015}.

In this work we forecast the ability of future gamma-ray observations of dSphs to constrain the velocity dependence of  dark matter annihilation.  Our forecasts rely on a set of Milky Way dSphs with $J$-factors measured in \cite{Boddy:2020}.  We consider both current $J$-factor uncertainties, as well as prospects for future improvements.  We generate mock data sets for the Cherenkov Telescope Array (CTA), the Fermi Gamma-Ray Space Telescope, and the All Sky Medium Energy Gamma-Ray Observatory (AMEGO).   For each observatory, we consider a baseline exposure as well as significantly enhanced exposures.  The mock data sets include realistic estimates of backgrounds.  Using these mock observations, we estimate the future improvements 
that will be needed in order to distinguish between 
different models of dark matter microphysics from the data.  We discuss possible sources of systematic error, and how these may impact future attempts to infer the dark matter velocity dependence from future dSph observations.  We note, though, that the main aim of this paper is {\it not} to produce the most accurate forecasts possible.  Indeed, making very realistic or precise forecasts for dSph observations is complicated by the fact that future constraints will depend to some degree on the intrinsic dSph properties, which are not very well constrained at present.  Rather, the main aim of this analysis is to highlight that information about the dark matter annihilation velocity-dependence is contained in the relative amplitude of annihilation signals from different dSphs, and that in principle, there is sufficient {\it statistical} information in future datasets to constrain this dependence.

The paper is organized as follows.  In \S\ref{sec:formalism} we introduce the formalism for modeling the velocity-dependent $J$-factors of dSphs; in \S\ref{sec:forecast_methods} we describe the $J$-factor constraints for a set of dSphs, and how we generate forecasts for the constraints on the dark matter annihilation velocity dependence.  Our results are presented in \S\ref{sec:results}, and we conclude in \S\ref{sec:discussion}.

\section{General formalism}
\label{sec:formalism}

We assume that dark matter is a real particle with 
an annihilation cross section given by 
\bea
\sigma v &=& (\sigma v)_0 \times  S(v/c),
\eea
where $v$ is the relative velocity between the 
dark matter particles, and $(\sigma v)_0$ is a 
constant which is independent of $v$.  The velocity 
dependence of the annihilation process is contained 
in $S(v/c)$, which we will assume takes the form 
$S(v/c) = (v/c)^n$.  We will consider several 
theoretically-motivated choices for $n$.
\begin{itemize}
\item{$n=0$ ({\it s-wave}): This is the standard case of 
velocity-independent annihilation.}
\item{$n=2$ ({\it p-wave}):
This case can arise in 
any scenario respecting minimal flavor violation (MFV) in which dark matter is a Majorana fermion 
which annihilates to a Standard Model (SM) 
fermion/anti-fermion pair (see, for 
example,~\cite{Kumar:2013iva}). In this case, annihilation 
from 
an $L=0$ state is chirality-suppressed, and annihilation 
from the $L=1$ state may thus dominate.  This case 
can also arise if dark matter is a fermion (Majorana or 
Dirac) which annihilates through an intermediate scalar 
in the $s$-channel.}
\item{$n=4$ ({\it d-wave}):
This case can arise in any 
scenario respecting MFV in which dark matter is a real 
scalar particle, which annihilates to a SM fermion/anti-fermion 
pair~\cite{Giacchino:2013bta,Toma:2013bka}.  
In this case, annihilation from the $L=0$ state is 
chirality-suppressed, while annihilation from the 
$L=1$ state is forbidden by symmetry of the 
wavefunction~\cite{Kumar:2013iva,Giacchino:2013bta,Toma:2013bka}.  
Annihilation from the $L=2$ state may thus dominate.}
\item{$n=-1$ ({\it Sommerfeld-enhanement in the 
Coulomb limit}):  This case 
can arise if dark matter annihilation is Sommerfeld-enhanced, 
and the particle mediating dark matter self-interaction is 
much lighter than the dark matter~\cite{ArkaniHamed:2008qn,Feng:2010zp}.}
\end{itemize}

The expected number  of photons with energies between $E_{\rm min}$ and $E_{\rm max}$ 
arising from dark matter annihilation in any 
astrophysical target can be written as~\cite{GeringerSameth:2011iw,Boddy:2018qur}
\bea
\label{eq:expected_counts}
N_{\rm exp} &=& 
\Phi_{PP} \times J(\Delta \Omega) \times 
(T A_{eff}),
\eea
where $T$ is the exposure time, $A_{eff}$ is the 
effective area, 
\bea
\Phi_{PP} &\equiv& \frac{(\sigma v)_0}{8\pi m_\chi^2} 
\int_{E_{\rm min}}^{E_{\rm max}} dE_\gamma ~ 
\frac{dN_\gamma}{dE_\gamma},
\eea
$m_\chi$ is the dark matter mass, and 
$dN_\gamma / dE_\gamma$ is the photon spectrum per 
annihilation. 
The integrated $J$-factor is given by
\bea
J (\Delta \Omega) &=& \int_{\Delta \Omega} d\Omega 
\int d\ell \int d^3 v_1 \int d^3 v_2 ~
f(\vec{r}, \vec{v}_1) f(\vec{r},\vec{v}_2) \times 
S(|\vec{v}_1 - \vec{v}_2|/c) ,
\eea
where $f(\vec{r}, \vec{v})$ is the dark matter 
velocity distribution, $\Delta \Omega$ is the solid 
angle, and $\ell = |\vec{\ell}|$ 
is the distance along the line of sight.  
If $\vec{D}$ is a vector from the observatory to the 
center of the dSph, then $\vec{r} = \vec{\ell} - 
\vec{D}$.

We thus see that $\Phi_{PP}$ depends only on the 
properties of the dark matter particle, while all of 
the dependence of the photon counts on the dark matter 
distribution in the target appears in the $J$-factor.  
But the $J$-factor also depends on $S(v/c)$.  For the 
case of $s$-wave dark matter annihilation ($S=1$),
the $J$-factor 
reduces to the usual expression 
$J(\Delta \Omega) = \int d\Omega~d\ell~\rho^2$.  But 
for a more general particle physics model, the $J$-factor 
of the target must be recomputed.

The form of the $J$-factor simplifies considerably 
for the case in which the dark matter velocity distribution 
depends on only two parameters, a scale density 
$\rho_s$ and a scale radius $r_s$.  One then finds 
that 
the only quantity one can write with units of velocity 
which depends on the relevant parameters is 
$4\pi G_N \rho_s r_s^2$.  The form of the $J$-factor 
simplifies even more if the dSph is reasonably 
far away ($D \gg r_s$), and the aperture of the observation 
covers the region where most dark matter annihilation occurs. 
The dependence of the integrated $J$-factor on the parameters is 
then determined by dimensional analysis, yielding~\cite{Boddy:2019wfg}  
\bea
J (\Delta \Omega) &\propto&  
\frac{\rho_s^2 r_s^3}{D^2} \left(4\pi G_N \rho_s r_s^2 
\right)^{n/2} ,
\label{eqn:JFactorDependence}
\eea
where the proportionality constant is independent of the 
halo parameters.

Given any ansatz for the form of the dark matter 
distribution, 
stellar data can be used to estimate the halo parameters, 
which in turn determine the $J$-factor for any choice of 
$S(v/c)$.  However, because the annihilation flux also depends on $\Phi_{PP}$, measurement of the flux from a single dSph will be insufficient to determine both $\Phi_{PP}$ and $n$.  On the other hand, if one considers the ratio of the fluxes between two dSphs with different velocity distributions, this ratio will be independent of $\Phi_{PP}$, but will depend on $n$ (and the velocity distributions). This implies that, if the halo parameters of several dSphs can be determined with sufficient precision from stellar data, with 
a sufficient exposure, it should be possible to determine 
$n$ from the relative photon counts from different dSphs.

For our analysis, we will consider the $J$-factors derived 
in~\cite{Boddy:2020} for 25 dSphs, assuming either 
$n=-1, 0, 2$, or $4$.  Following~\cite{Pace:2018tin}, the analysis of \cite{Boddy:2020}
assumed an NFW profile, and estimated $\rho_s$ and $r_s$ for 
each dSph from stellar data.  The dark matter 
velocity distribution was then determined from the density 
distribution using the Eddington inversion method~\cite{Widrow:2000fv}, 
following~\cite{Boddy:2019wfg}; this determines the 
proportionality constant in eqn.~\ref{eqn:JFactorDependence} 
for each choice of $n$.  Note that, although this overall 
proportionality constant affects the normalization of the dark matter signal from a dSph, it does not affect 
one's ability to determine the velocity-dependence of a 
detected signal {\it at fixed signal flux}, which depends on the {\it relative} flux between different dSphs.  
We discuss the estimation of $J$-factors and forecasts for future $J$-factor constraints in more detail in \S\ref{sec:J_forecasts}.
We will also consider $J$-factors which we 
derive using a modified version of the approach used in~\cite{Pace:2018tin,Boddy:2020}, in which the stellar data is supplemented with a cosmological prior derived from numerical simulations.

\section{Forecasting future constraints on the dark matter annihilation velocity-dependence}
\label{sec:forecast_methods}

\subsection{The likelihood for photon counts from dSphs}

We consider here the case of \textit{unresolved} observations of the annihilation signal in dSphs.  Because the observations are unresolved, we define our observable to be the measured photon counts in an aperture around each dSph.  Since the dark matter annihilation signal is expected to be localized in a small region centered on each dSph and because the beam size of gamma-ray telescopes is typically large compared to these regions, assuming that the annihilation signal is unresolved is reasonable.  For high-resolution observations, such as with CTA, it maybe be possible to improve constraints on the velocity dependence by using the angular dependence of the signal \cite{Boddy:2019wfg}.

For a set of $N_D$ dSphs, we define a $N_D$-dimensional data vector, $\vec{d}$, that represents the photon counts in the aperture around each dSph.  The observed data is the sum of signal photons and background photons:
\begin{eqnarray}
\vec{d} = \vec{s} + \vec{b},
\end{eqnarray}
where the $N_D$-dimensional vectors $\vec{s}$ and $\vec{b}$ represent the photon counts from signal and backgrounds, respectively.  We represent the probability distribution functions (PDF) describing  $\vec{s}$ and $\vec{b}$ as $P_S(\vec{s})$ and $P_B(\vec{b})$, respectively.  We will discuss $P_B(\vec{b})$ in more detail in \S\ref{sec:background_modeling}.

We assume that the dark matter signal, $\vec{s}$, is Poisson distributed.   The expectation value of the signal for the $i$th dSph, $\langle s_i \rangle$, is given by $\langle s_i \rangle = N_{\rm exp}$ (Eq.~\ref{eq:expected_counts}).  The signal PDF is then
\begin{equation}
\label{eq:signal_poisson}
    P_S(s_i | J_i, \Phi_{PP}) = \frac{\langle s_i \rangle^{s_i} e^{-\langle s_i \rangle}}{s_i !},
\end{equation}
where $\langle s_i \rangle$ depends on the dSph's $J$-factor, $J_i$, the particle physics factor $\Phi_{PP}$, and the exposure.  We remind the reader that the $J$-factor in turn depends on the velocity dependence, $n$.  Since the observed sky signal is the sum of signal and backgrounds, the total data likelihood is given by a convolution of the signal and background distributions:
\begin{eqnarray}
P_{D}(d_i | J_i, \Phi_{PP}) = \sum_{j=0}^{d_i}P_S(j 
| J_i, \Phi_{PP}) P_B(d_i-j).
\end{eqnarray}

Ultimately, we are interested in constraining the velocity dependence of the dark matter annihilation (i.e. $n$), rather than the $J$-factors themselves.  Marginalizing over the $J$-factor PDF we have
\begin{eqnarray}
P_{D}(d_i | \Phi_{PP}, n) = \int dJ_i \, P_{D}(d_i | J_i, \Phi_{PP})  P_J(J_i | n),
\end{eqnarray}
where $P_J(J_i|n)$ is the prior on the $J$-factor of the $i$-th dSph, which we will discuss in more detail in \S\ref{sec:J_forecasts}.  

Assuming the dSphs are far enough apart on the sky that they can be treated as statistically independent, we write the total likelihood for all dSphs as
\begin{equation}
\label{eq:likelihood}
\mathcal{L} \equiv P_D(\vec{d} |  \Phi_{PP}, n) = \prod_i^{N_D} P_D(d_i | \Phi_{PP}, n).
\end{equation}
We adopt flat priors on $\Phi_{PP}$ and $n$ so that the posterior on $\Phi_{PP}$ and $n$ is simply proportional to this likelihood.  The purpose of our analysis is to determine whether (future) observations can distinguish between different models for the velocity dependence of the dark matter annihilation cross section.  In \S\ref{sec:forecasts} we describe how the likelihood introduced above can be applied to mock data to make such forecasts.

\subsection{Background modeling}
\label{sec:background_modeling}

We will make forecasts for future observations in three energy ranges: (1) $1-100\gev$, (2) $1-200\tev$, and (3) $1\mev-1\gev$.  In each case, we will take different approaches to estimating the PDF describing photon backgrounds, $P_B(\vec{b})$.  

Our analysis at $1-100\gev$ is modelled after Fermi observations.  In this case, we will use the Fermi maps themselves to estimate the backgrounds.  This can be done by defining a large number of background 
sky regions which are of the same size as the 
signal aperture, but displaced 
slightly from the dSph; the histogram of photon counts 
in these backgrounds regions forms our estimate of the background PDF for 
that dSph.  This procedure has been applied in \cite{GeringerSameth:2011iw,Boddy:2018qur,Boddy:2019kuw}, for 
example, and we will use the background PDFs obtained in
Ref.~\cite{Boddy:2019kuw}.  
Note that these PDFs can be highly non-Poissonian, owing largely to the complicated morphology of the diffuse galactic backgrounds.  For our baseline analysis, we adopt an exposure corresponding to roughly 10 years of observation time with Fermi, i.e. the data set used in \cite{Boddy:2019kuw}; the exact exposure values assumed for each dSph are given in the appendix of \cite{Boddy:2019kuw}.  We will also consider a future Fermi-like data set that has a factor of five larger exposure, which could be obtained by increasing the observation time and/or collecting area relative to Fermi.

Our analysis at higher photon energies is tailored to CTA-like observations.  At energies $E \gtrsim 100 \,{\rm GeV}$ and for detectors like CTA, the dominant background is cosmic rays that have been misclassified as gamma-rays
(the so-called residual background).  
Since this background is close to isotropic, we can ignore the sky positions of the dSphs, and obtain an accurate estimate of the backgrounds by using the estimated spectrum of these misclassifications. Since the effective area of CTA is both maximal and approximately constant for photon energies $1\,{\rm TeV} \lesssim E \lesssim 200\,{\rm TeV}$, we assume this energy range in our analysis. We adopt the reported background flux for CTA south.\footnote{\url{https://www.cta-observatory.org/science/cta-performance/}} We assume an exposure time (for each dSph) of 20 hours, an effective area of $4\times 10^6\,{\rm m}^2$, and an aperture of radius $0.5^{\circ}$.  This aperture size matches that used in the analysis of \cite{Boddy:2019kuw}; significantly smaller apertures would remove signal flux, while much larger apertures would significantly increase the backgrounds.  Since the CTA beam size at these energies is roughly $0.03^{\circ}$, more information about the velocity dependence of the dark matter annihilation 
cross section 
could be obtained by considering the angular dependence of the signal rather than the total flux in an aperture.  For the present analysis, though, we ignore the angular dependence, so our constraints can be viewed as conservative. 

We also consider the energy range 
$E \lesssim 1~\gev$, for which future MeV-range 
gamma-ray telescopes, such as e-ASTROGAM, AMEGO and 
APT, can conduct a similar search for 
photons from dSphs.  
For this energy range, one would expect the astrophysical 
background to be anisotropic.  Unfortunately, however, we do not have a 
data-driven background estimate for individual dSphs over this energy range.  Instead, as a benchmark, we will use a fit to the 
isotropic background seen by COMPTEL (0.8-30 MeV) and 
EGRET (30 MeV - 10 GeV).  This fit is given by~\cite{Boddy:2015efa}
\begin{eqnarray}
\frac{d^2 \Phi}{dE d\Omega} &=& 
2.74 \times 10^{-3} \left(\frac{E}{\rm{MeV}} \right)^{-2.0}
\rm{cm}^{-2} \rm{s}^{-1} \rm{MeV}^{-1} \rm{sr}^{-1} .
\end{eqnarray} 
Integrating this fit over the desired energy range provides a rough estimate of the expected background flux.  We tailor our low-energy forecasts to AMEGO-like observations, assuming an energy range of $1\,{\rm MeV} < E < 1\,{\rm GeV}$, a baseline exposure time of one year, an effective area of $800\,{\rm cm}^2$, and a beam size of $2.5^{\circ}$ \cite{AMEGO}.

\subsection{$J$-factors and their uncertainties}
\label{sec:J_forecasts}

As mentioned previously, \cite{Boddy:2020} constrained $P_J(J_i|n)$ by running fits to stellar velocity data.  
The full details are described in \cite{Pace:2018tin}.
Briefly, the Spherical Jeans equations are solved for the radial velocity dispersion which is  projected into the line-of-sight direction to directly compare to stellar velocity data \cite{Strigari2008ApJ...678..614S, Bonnivard2015MNRAS.453..849B, Geringer-Sameth2015ApJ...801...74G}. 
The Spherical Jeans equations are solved assuming an NFW profile for the dark matter distribution, a Plummer profile for the stellar distribution, and a constant stellar anisotropy. 

Several of these model assumptions are known to be broken in reality.  For instance, simulations suggest that the dark matter halos hosting dSphs are likely triaxial \citep[e.g.][]{Munoz:2011}, the stellar distribution may not be described by a Plummer profile, and the velocity anisotropy may vary with radius~\cite{Bonnivard:2015}.  As discussed in \cite{Bonnivard:2015}, these incorrect assumptions when modeling the stellar data can result in biased $J$-factor constraints, although for current data the biases appear to be fairly small~\cite{Sanders:2016, Pace:2018tin}.  In principle, sources of bias can be eliminated by adopting more flexible forms for the assumed profiles of e.g. the dark matter or the stellar velocity anisotropy~\cite{Bonnivard:2015}.  However, this will come at the cost of increased $J$-factor uncertainties.  We discuss our approach to dealing with systematic errors below.

For the purposes of this analysis, we will assume that the posteriors on the $J$-factor for each dSph is described by a Gaussian:
\begin{eqnarray}
P_J(J_i|n) \propto \exp \left[ -\frac{(J_i - \mu_J)^2}{2\sigma_{J,i}^2}\right],
\end{eqnarray}
where $\mu_J$ and $\sigma_J$ are the mean and standard deviations computed from the posterior samples generated in \cite{Boddy:2020}.
Assuming Gaussianity is useful partly because it allows us to trivially make forecasts for future data by appropriately reducing $\sigma_{J,i}$.  For current data, the $J$-factor posteriors for individual dSphs can be significantly non-Gaussian for reasons that we discuss in \S\ref{sec:jfactor_uncertainty}.  However, we show below that approximating the individual $J$-factor posteriors as Gaussians does not lead to significant error in the combined constraints from all dSphs.  Furthermore, the Gaussian approximation is likely to become more accurate for individual dSphs as stellar velocity constraints improve.

Future stellar observations will improve $J$-factor constraints by measuring velocities for fainter stars. To make projections for the estimated $J$-factor uncertainty with different stellar magnitude cuts, we first make forecasts for how the number of stars observed in the dSphs will increase with future observations.
We estimated the number of stars at different magnitudes in each dSph by drawing stars from an  initial mass function \cite{Chabrier2001ApJ...554.1274C} with a metallicity of [Fe/H]=-2.2 and an age of 12.5~Gyr \citep{Bressan2012MNRAS.427..127B}. 
Taking the dSph absolute magnitudes compiled in \cite{Pace:2018tin} and assuming a mass-to-light ratio of 2 we performed 1000 simulations with the Ultra-faint Galaxy Likelihood
(\texttt{ugali}) software toolkit\footnote{\url{https://github.com/DarkEnergySurvey/ugali}} \cite{Bechtol2015ApJ...807...50B, DrlicaWagner2015ApJ...813..109D} to estimate the number of stars at a given magnitude. 
At each magnitude limit we assumed that all stars brighter than this are observed. 
We assume a limiting magnitude of $r_{\rm DECam}<23.5$ which is expected for future 30m class telescopes with multi-object spectrographs such as GMT/GMACS \citep{2018SPIE10702E..1XD} and E-ELT/MOSAIC \citep{2015arXiv150104726E}.  
Finally, we assume that the $J$-factor uncertainty scales according to $\sigma_{\rm future}(J) = \sigma_{\rm current}(J)\sqrt{N_{\rm current}/N_{\rm future}}$, where $N$ is the number of stars in a dSph and the subscripts indicate current or forecast observations.

So far, we have only accounted for the statistical uncertainty on the $J$-factors, which can be reduced by observing more stars.  As mentioned above, we must also contend with systematic errors due to, for instance, incorrect modeling assumptions and potential unresolved binary stars.  Some of these sources of systematic error are likely to be reduced in the future.  For instance, high resolution simulations may be used to provide useful priors on the degree of dSph triaxiality, and the impact of baryons on the dark matter profile.  Similarly, increasing the sample of line-of-sight velocities or future tangential velocity measurements with {\it Gaia}, or other space based astrometry (e.g., the  {\it James Webb Space Telescope }), may be able to reduce uncertainty on, e.g., the degree of stellar velocity anisotropy. In addition, multi-epoch velocity data can identify unresolved binary stars  \citep[e.g.,][]{Martinez2011ApJ...738...55M, Kirby2017ApJ...838...83K}. However, there are also systematic errors, such as the dark matter velocity distribution, that may be very difficult to reduce, even with future data and improved simulations.

We adopt a simple and conservative prescription for including systematic uncertainties on the $J$-factors in our analysis.  Several authors \cite{Bonnivard:2015, Hayashi:2016, Sanders:2016} find that the impact of allowing triaxial dark matter profiles can change the inferred $J$-factors by factors of a few.  Similarly, \cite{Bonnivard:2015} find that assuming the incorrect stellar distribution or velocity anisotropy can bias the $J$-factors by factors of a few.   To roughly account for systematic errors, then, we perform an analysis where the uncertainties on $\log_{10} J$ for all dSphs are increased by $0.5$, corresponding to a factor of roughly three uncertainty.  This approach assumes that future analyses adopt sufficiently flexible profile models so that unbiased constraints on the $J$-factors can be obtained, albeit with higher uncertainties. 

\subsection{Imposing a prior on the $r_s$-$\rho_s$ relation}
\label{sec:Jwithprior}

The analysis of \cite{Boddy:2020} does not impose any informative prior on the relationship between $r_s$ and $\rho_s$ when fitting to the stellar velocity data.  As we discuss in \S\ref{sec:jfactor_uncertainty}, strong parameter degeneracies degrade the precision of the resultant $J$-factor constraints.  Numerical simulations predict that $r_s$ and $\rho_s$ are related for cold dark matter halos, and by imposing a prior on this relationship we can change the inference of the $J$-factors and potentially improve the $J$-factor precision.  A similar point has recently been made by \cite{Ando:2021}. 

Following \cite{Martinez:2009jh,Boddy:2017vpe}, we adopt a Gaussian prior with mean
\begin{eqnarray}
\langle \log_{10}(r_{\rm max}/{\rm kpc})  \rangle &=& 1.35\log_{10}\left(V_{\rm max}/({\rm km}/{\rm s}) \right)-1.75,
\end{eqnarray}
and standard deviation
\begin{eqnarray}
\sigma (\log_{10}(r_{\rm max}/{\rm kpc})) &=& 0.22,
\end{eqnarray}
 where $r_{\rm max} = 2.16 r_s$ and $V_{\rm max} = 0.465\sqrt{4\pi G \rho_s r_s^2}$.  
This relation~\cite{Martinez:2009jh} was found from a fit to subhalos in the Aquarius 
 simulations~\cite{Springel:2008cc}.
 
We present an alternative derivation of 
the dSph velocity-dependent $J$-factors, using the same 
posterior samples as in Ref.~\cite{Boddy:2020}, but with an 
additional weighting by the cosmological prior given above.  
The resultant $J$-factor constraints are presented in Appendix~\ref{app:J_constraints}.
Below, we will present results utilizing $J$-factors derived 
both with and without this $J$-factor prior. 

\section{Results}
\label{sec:results}

\begin{figure}
    \centering
    \includegraphics[scale=0.5]{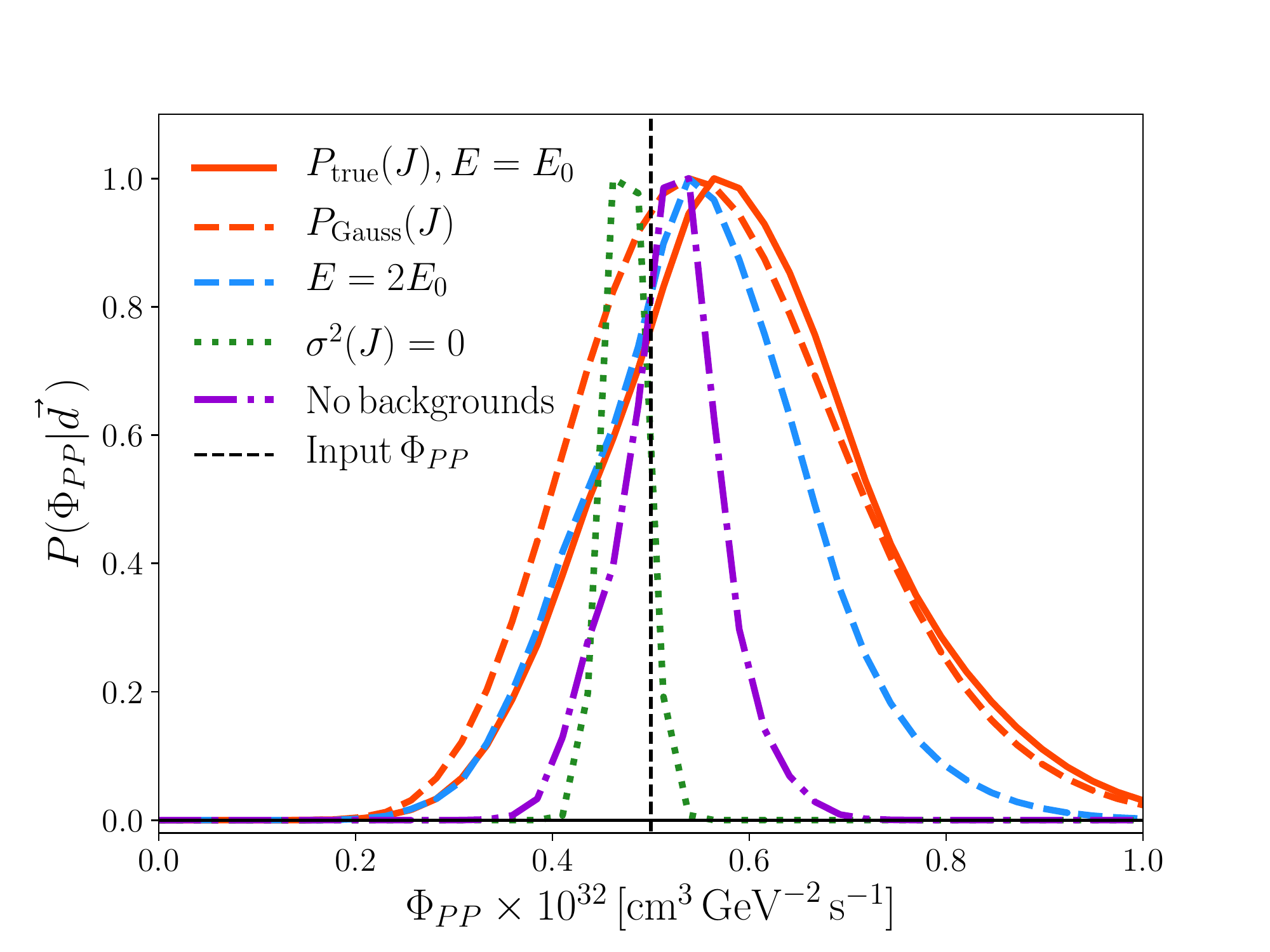}
    \caption{Recovered posterior on $\Phi_{PP}$ when the mock data are generated assuming $s$-wave annihilation and CTA-like observations.  We assume $s$-wave annihilation when analyzing the mock data, so the posterior should recover the input value of $\Phi_{PP}$, shown with the vertical dashed line; this value is set larger than current limits for illustrative purposes.   The red curves show the posterior for our baseline model assumptions, computed using the true posteriors on the dSph $J$-factors (solid) and a Gaussian approximation to the $J$-factor posteriors (dashed).  The blue dashed curve shows the impact of increasing the exposure by a factor of two, which significantly improves our ability to constrain $\Phi_{\rm PP}$.  We also show the impact of removing $J$-factor uncertainty (green dotted) and backgrounds (purple dot-dashed).  We see that $J$-factor uncertainty significantly degrades our ability to recover $\Phi_{PP}$.  The assumed energy range is $1\,{\rm TeV} < E < 200\,{\rm TeV}$.}
    \label{fig:posterior}
\end{figure}

\subsection{Generating and analyzing mock data}
\label{sec:phipp_recovery}

We generate mock data as follows.  First, we assign a true $J$-factor to every dSph.  The true $J$-factors are set to $\mu_J$, i.e. the mean $J$ values from the analysis of stellar data described in \S\ref{sec:J_forecasts}.  We then randomly draw from the Poisson distributions in Eq.~\ref{eq:signal_poisson} to assign a mock dark matter annihilation signal to each dSph.  We next draw from the background distributions for each dSph to assign them mock background photon counts.  The combined signal and background counts for each dSph represent a mock data set that we can analyze using the likelihood defined in Eq.~\ref{eq:likelihood}.

In Fig.~\ref{fig:posterior} we show the posteriors on $\Phi_{PP}$ that we obtain from our analyses of mock CTA data, computed by evaluating Eq.~\ref{eq:likelihood} across a grid of $\Phi_{PP}$ values.
Each curve represents the posterior, $P(\Phi_{PP} | \vec{d})$, obtained from combining the constraints across all dSphs for a different realization of mock data.  For this figure we assume $s$-wave annihilation, with $\Phi_{PP} = 5\times 10^{-33}\, \mathrm{cm^3\, s^{-1}\, GeV^{-2}}$ (shown with the vertical black dashed line).

The red solid curve shows the recovered constraint on $\Phi_{PP}$ assuming an exposure of $E_0 = 20\,{\rm hrs}$ (per dSph) and using the true (non-Gaussian) $J$-factor PDFs from \cite{Boddy:2020}.  As expected, we recover the input value of $\Phi_{PP}$ to within the uncertainties.  For comparison, the dashed red curve shows the results of analyzing the same mock data set, but approximating the $J$-factor PDFs with Gaussians of the same variance.  This approximation introduces some error in the posterior, but it is small compared to the uncertainty on $\Phi_{PP}$.  

The blue dashed curve shows the results of increasing the exposure by a factor of two, computed using the true $J$-factor posteriors.  In this case, the width of the posterior is reduced.  The green dotted curve represents the case where we know the $J$-factor exactly, while the purple dot-dashed curve represents the case with no background photons.  For the assumed value of $\Phi_{PP}$, uncertainty on the $J$-factors dominates over uncertainty from the backgrounds.  At lower $\Phi_{PP}$ or lower exposure, though, uncertainty contributed by the backgrounds can become significant.  Note that we expect scatter between the different curves in this plot, as each one corresponds to a different random realization of the mock data (except the two red curves, which represent analyses of the same mock data).

\subsection{Ability of future dSph observations to constrain the dark matter annihilation velocity dependence}
\label{sec:forecasts}

We now forecast the ability of future observations of dSphs to constrain the velocity dependence of dark matter annihilation by analyzing mock data sets generated as described above.  We generate a mock data set, assuming a true model (that is, a
choice of $n$ and $\Phi_{PP}$), 
along with a choice of exposure and choice of $J$-factor 
uncertainties (i.e., either current uncertainties or forecast uncertainties for fainter stellar samples).
Given this mock data set, we maximize the likelihood 
over $\Phi_{PP}$, assuming either the value of $n$ used to generate the mock data or 
an alternative choice.  

For two models (model 1 and model 2), the difference in the maximum 
likelihoods, $\Delta \ln \mathcal{L}_{\rm max} = \mathcal{L}_{\rm max,1} - \mathcal{L}_{\rm max,2}$, 
is related to our ability to
reject model 2 in favor of model 1, based on the data.  For instance, two common criteria for model selection are the Akaike information criterion (AIC) and the Bayesian information criterion (BIC), both of which are related to the $\Delta \ln \mathcal{L}_{\rm max}$ between models \citep[e.g., ][]{Elements}.
The BIC, for example, is given by 
\begin{eqnarray}
\mathrm{BIC} = k \ln N_D  - 2\ln \mathcal{L}_{\rm max},
\end{eqnarray}
where $k$ is the number of free parameters in the model (i.e. $k=1$ when $\Phi_{PP}$ is varied, and $k=0$ for the null model that has no dark matter signal), and $N_D$ is again the number of dSphs.   Given a set of models, it can be shown that under certain approximations, the posterior probability of model $i$ is proportional to $\exp [-\mathrm{BIC}_i/2]$ \cite{Elements}.  
Since here $k\ln N_d$ is small, when comparing two fits to the data with different values of $n$, if $\Delta \mathcal{L}_{\rm max} \gg 1$ the model with the larger maximum likelihood will be strongly favored over the other model.  In the present circumstances, the difference between e.g. the AIC and BIC will be small, since we are most interested in cases where $\ln \mathcal{L}_{\rm max}$ is large.  Below, we will report the $\Delta \ln \mathcal{L}_{\rm max}$ between the true model (i.e. the one used to generate the data) and an alternate model.

\begin{figure}
    \centering
    \includegraphics[scale=0.55]{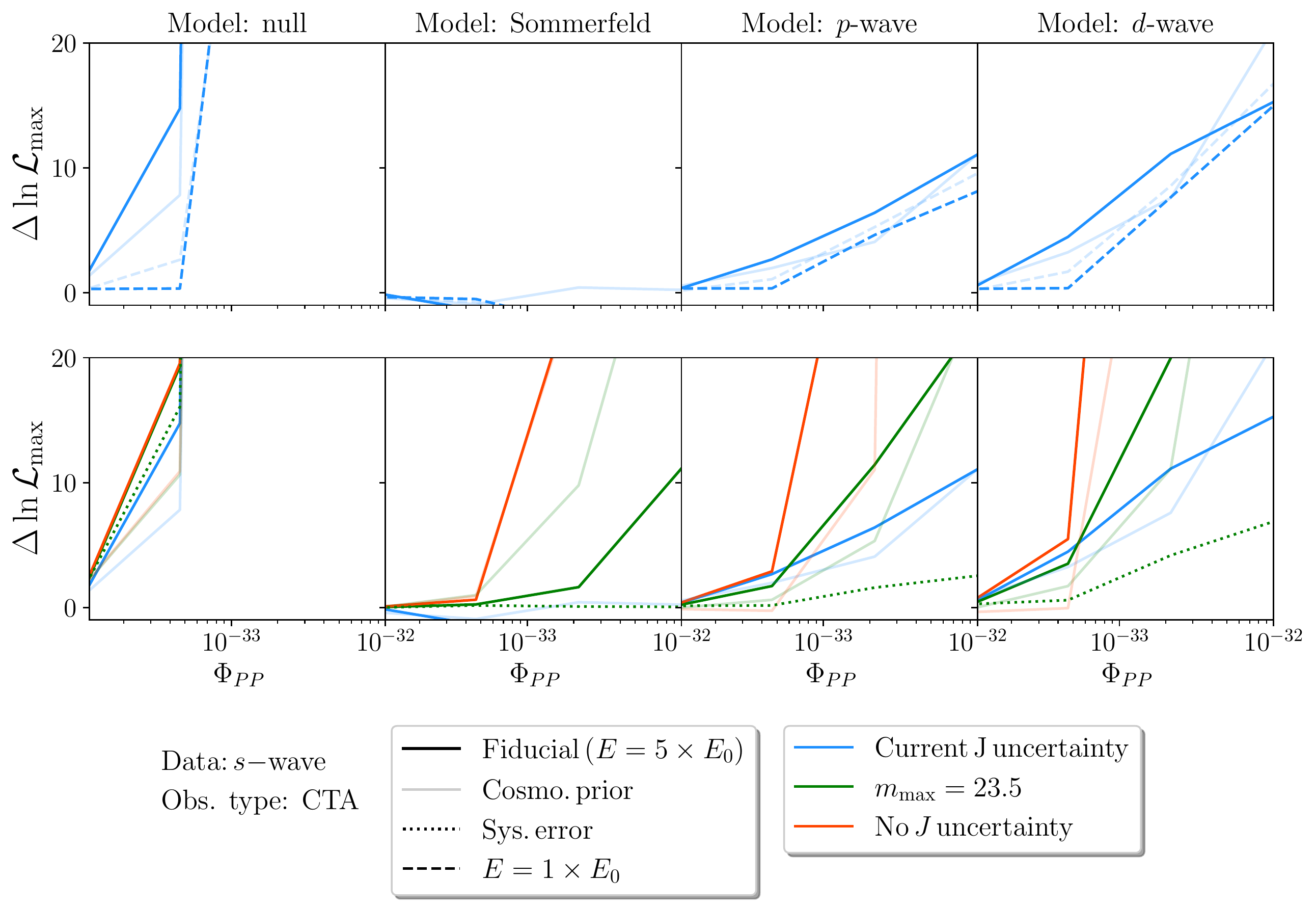}
    \caption{Results of the mock data analysis.  We plot the $\Delta \ln \mathcal{L}_{\rm max}$ between the model assuming the true annihilation velocity dependence 
    --- in this case, $s$-wave --- and the alternate model
    labeled atop each column.  The mock data are generated assuming CTA-like observations.   
    Our fiducial result (solid curve) assumes an exposure of $5 E_0$, no cosmological prior, and no systematic $J$-factor error.   The faded curves show the impact of imposing the $r_s$-$\rho_s$ prior, dotted curves show the impact of including systematic $J$-factor error, and the dashed curves shows the impact of reducing the exposure to $E_0$.   The units of $\Phi_{PP}$ are ${\cm^3} 
    {\rm s}^{-1} {\rm GeV}^{-2} $, and the assumed energy range is $1\,{\rm TeV} < E < 200\,{\rm TeV}$.}
    \label{fig:deltaloglike_CTA_sdefault}
\end{figure}

Fig.~\ref{fig:deltaloglike_CTA_sdefault} shows the ability of future observations of dSphs with a CTA-like experiment to distinguish 
different alternative velocity-dependent annihilation models, 
assuming the true model is $s$-wave ($n=0$).  
On the $x$-axis we plot the value of $\Phi_{PP}$ used to generate the mock data set, and on the $y$-axis we plot $\Delta \ln \mathcal{L}_{\rm max}$ for $n=-1$ (second column), $n=2$ (third column) and 
$n=4$ (fourth column).  
The first column of Fig.~\ref{fig:deltaloglike_CTA_sdefault} represents the $\Delta \ln \mathcal{L}_{\rm max}$ between the true model and the model with $\Phi_{PP} = 0$ (i.e. no dark matter).  
For all panels, solid lines are used for analyses with an exposure $E_0 = 20\,{\rm hours}$, while the dashed lines 
are used for an exposure of $5E_0$.
Blue lines are used for analyses in which 
the velocity-dependent $J$-factors and their uncertainties are 
as found given in~\cite{Boddy:2020}.  Green lines are used for analyses in which the 
$J$-factor uncertainties are reduced, based on an estimate of what precision might 
be possible with a future survey with a magnitude limit of 23.5 (see \S\ref{sec:J_forecasts}).  Red lines correspond to the most optimistic case, in which the uncertainties in 
the $J$-factors are negligible.  Finally, the translucent lines correspond to 
analyses in which the $J$-factors and their uncertainties are derived using the cosmological prior 
described above.  
We note that, since we are analyzing simulated realizations of the data, we expect some scatter in the various curves with variance of order $\sigma^2(\Delta \ln \mathcal{L}_{\rm max}) \sim 1$.

Fig.~\ref{fig:deltaloglike_CTA_sdefault} also shows the impact of our assumed $\sigma(\log_{10} J) = 0.5$ systematic error (dotted curves).  It is clear that this level of systematic uncertainty significantly degrades the constraints.  This level of uncertainty corresponds to a (likely conservative) estimate of systematic uncertainties in current data.  Our analysis therefore provides additional motivation for reducing systematic uncertainties associated with 
$J$-factor measurements of dSphs.

\begin{figure}
    \centering
    \includegraphics[scale=0.55]{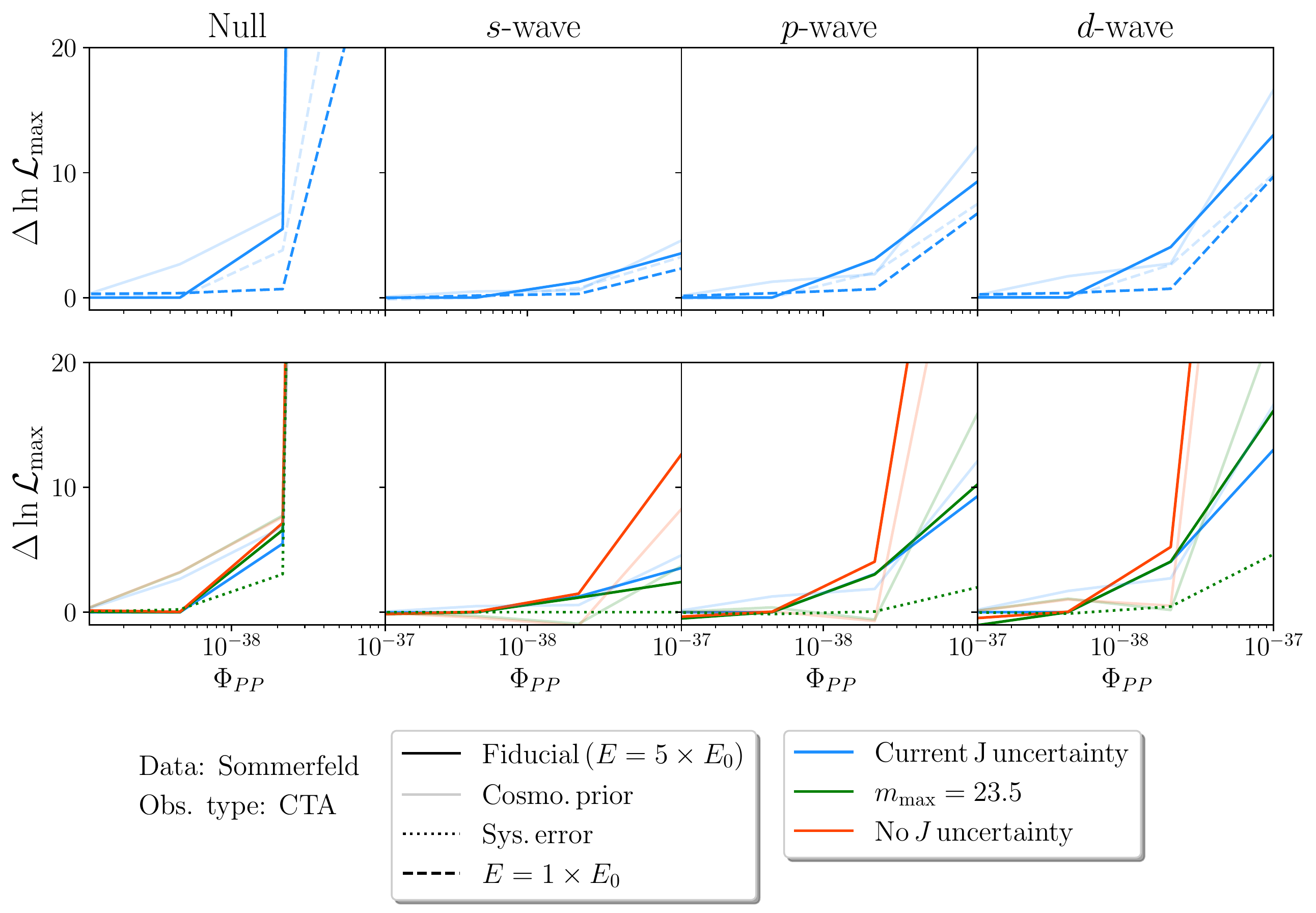}
    \caption{Same as Fig.~\ref{fig:deltaloglike_CTA_sdefault}, but for mock data generated assuming Sommerfeld velocity dependence.  The units of $\Phi_{PP}$ are $\cm^3 {\rm s}^{-1} {\rm GeV}^{-2} $.}
    \label{fig:deltaloglike_CTA_somdefault}
\end{figure}

\begin{figure}
    \centering
    \includegraphics[scale=0.55]{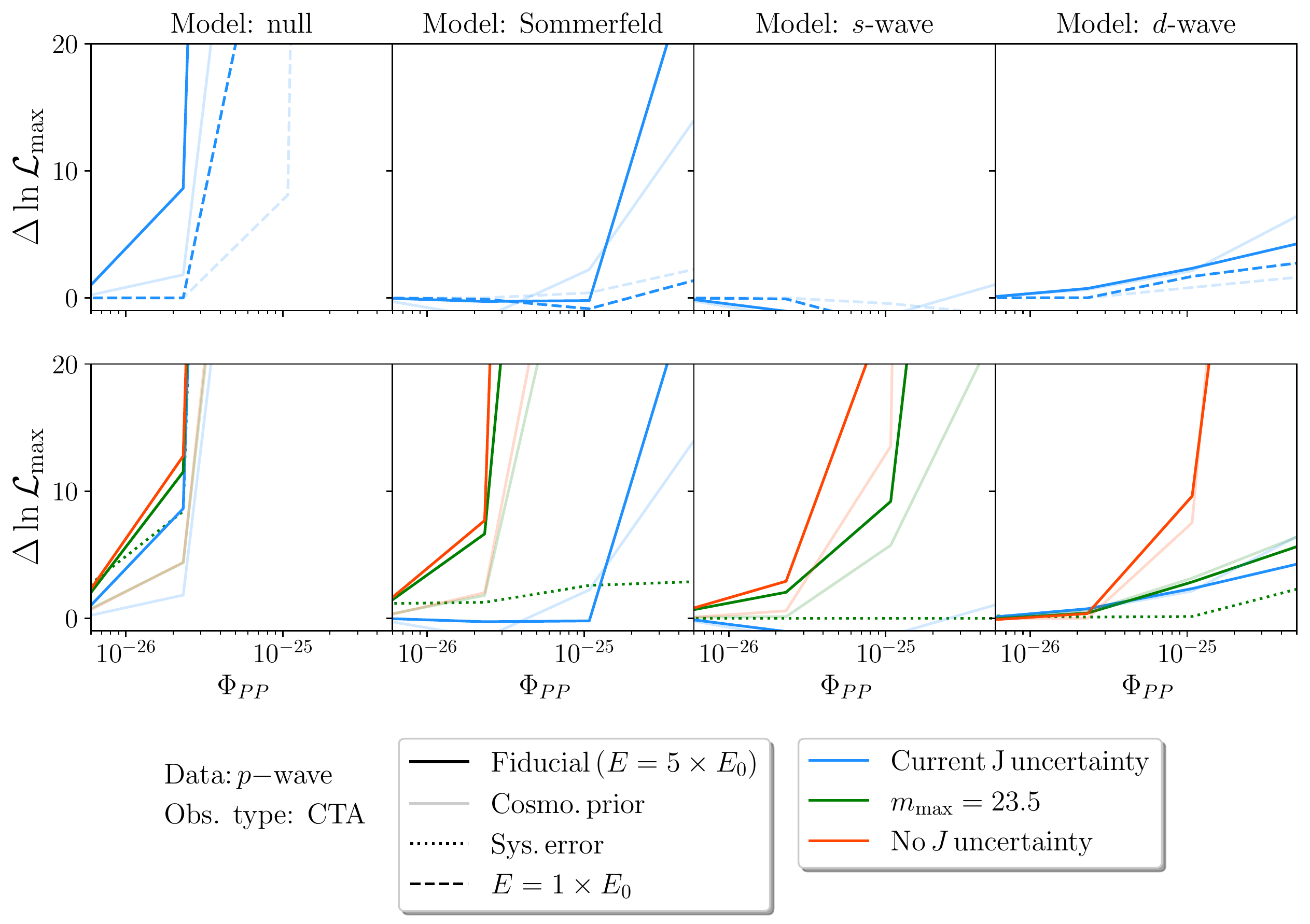}
    \caption{Same as Fig.~\ref{fig:deltaloglike_CTA_sdefault}, but for mock data generated assuming $p$-wave velocity dependence.  The units of $\Phi_{PP}$ are $\cm^3 
    {\rm s}^{-1} {\rm GeV}^{-2} $.}
    \label{fig:deltaloglike_CTA_pdefault}
\end{figure}

Fig.~\ref{fig:deltaloglike_CTA_somdefault} and Fig.~\ref{fig:deltaloglike_CTA_pdefault} are similar 
to Fig.~\ref{fig:deltaloglike_CTA_sdefault}, except that they use mock data generated assuming
Sommerfeld and $p$-wave annihilation, respectively.   In Appendix~\ref{app:alternate_results} we show the corresponding results for our Fermi and AMEGO forecasts.

We find that for sufficiently high $\Phi_{PP}$, different models of the velocity-dependence can be distinguished at high significance.  In general, the significance with which different velocity-dependence models can be distinguished is lower than that with which we can rule out the null model (i.e.~no dark matter).   This is sensible: we must be able to detect the dark matter signal before we can determine the velocity dependence of the dark matter annihilation 
cross setion.  We find that increasing the exposure time and decreasing 
the $J$-factor uncertainties improves the sensitivity significantly at high $\Phi_{PP}$.  

One perhaps surprising feature of Fig.~\ref{fig:deltaloglike_CTA_sdefault} is that when the data are generated with the $s$-wave model, we cannot rule out Sommerfeld annihilation for current $J$-factor uncertainty levels, regardless of how large $\Phi_{PP}$ is.  A second surprising feature of Fig.~\ref{fig:deltaloglike_CTA_sdefault} is that imposing a prior on the $r_s$-$\rho_s$ relationship does not necessarily help improve our ability to distinguish between different models for the velocity dependence.  Both of these features are connected to the dSph $J$-factors and their uncertainties, which we now consider in more detail.

\subsection{The impact of $J$-factor uncertainty}
\label{sec:jfactor_uncertainty}

The stellar velocity data used to constrain the $J$-factor effectively probes the circular velocity in the dSph, $V_{\rm circ}(r)$, at some radial distance, $r$, from the halo center.  The analysis in~\cite{Boddy:2020} assumed an NFW density profile for the dark matter, given by 
\begin{eqnarray}
\rho(r) = \frac{\rho_s}{(r/r_s)(1+ r/r_s)^2}.
\end{eqnarray}
Relating $V_{\rm circ}$ to the enclosed mass, $M(r)$, via $V_{\rm circ} \propto \sqrt{G_N M(r)/r}$ we have for $r \ll r_s$
\begin{eqnarray}
V_{\rm circ}(r) \propto \sqrt{G_N \rho_s r_s r},
\end{eqnarray}
while for $r \gg r_s$, we have 
\begin{eqnarray}
V_{\rm circ}(r) \propto \sqrt{G_N \rho_s r_s^3 \ln r / r}.
\end{eqnarray}
In each case, the $V_{\rm circ}$ measurements constrain some degenerate combination of $r_s$ and $\rho_s$, meaning that the constraints on the $J$-factor will be very weak.   If stars can be measured spanning a wide range of $r$, the different degeneracies in these two limits could be broken.  

\begin{figure}
    \centering
    \includegraphics[width=0.47\columnwidth]{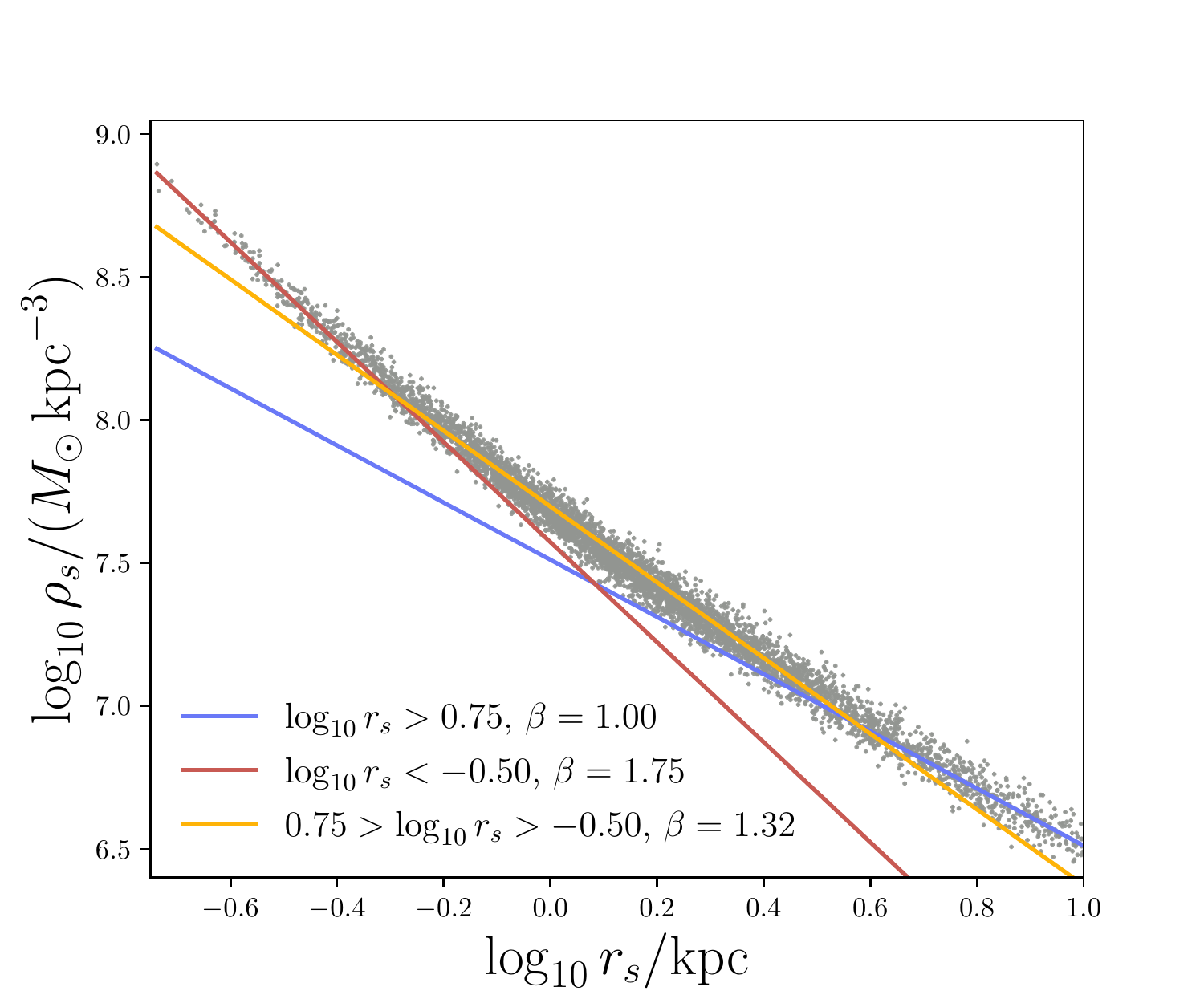}
    \includegraphics[width=0.47\columnwidth]{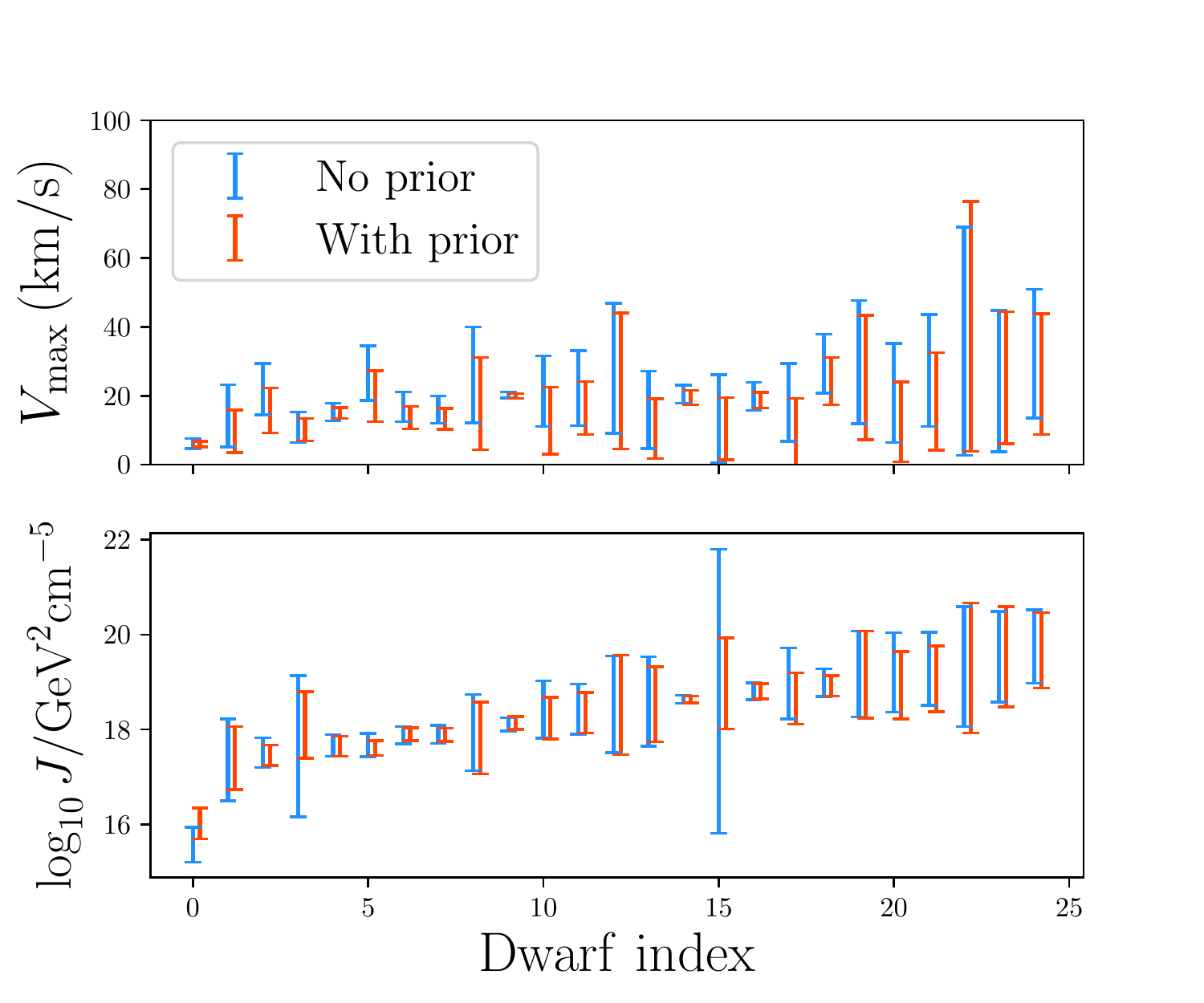}
    \caption{Left: degeneracy in the $r_s-\rho_s$ plane for the dSph Draco. The analysis of stellar velocity data leads to a degeneracy corresponding roughly to $\rho_s\propto r_s^{-\beta}$.  We include power law fits for three subsets of the data. 
    Right: constraints on $V_{\rm max}$ (top) and $\log_{10} J$ (bottom) assuming $s$-wave annihilation.  
    The dSphs are ordered by their $s$-wave $J$-factors.
    We show results with and without the imposition of the cosmological prior discussed in \S\ref{sec:Jwithprior}.  Given the large errorbars on $V_{\rm max}$ with current data, the measurements for different dSphs are roughly consistent.  We provide the $J$-factors assuming different velocity-dependence of the dark matter annihilation in Appendix~\ref{app:J_constraints}.}
    \label{fig:rhor}
\end{figure}

In practice, however, we find that for current dSph measurements, $r_s$ and $\rho_s$ remain quite degenerate with a degeneracy direction between these two limits, roughly $\rho_s \propto r_s^{-\beta}$ with $\beta \sim 1.3$.  This degeneracy is shown (for  Draco) in Fig.~\ref{fig:rhor} (left panel), in which we plot the values of 
$\rho_s$ and $r_s$ inferred from each the MCMC chains used 
in Ref.~\cite{Pace:2018tin}.  Using Eq.~\ref{eqn:JFactorDependence}, this translates into a degeneracy between $J$ and $r_s$ given by
\begin{eqnarray}
J \propto r_s^{3-2\beta+n(1-\beta/2)}.
\end{eqnarray}
Notably, since $3-2\beta > 0$ and $1-\beta/2 > 0$, 
the variation of $J$ with $r_s$ becomes steeper for larger values of $n$.  Since $r_s$ is only weakly constrained by the data, this means that the fractional $J$-factor uncertainty increases significantly for large $n$.

Furthermore, 
the values of $V_{\rm max} = V_{\rm circ}(r_{\rm max})$ (which is the maximum circular velocity) inferred from current data 
for the different dSphs are all very similar to each 
other, within uncertainties.  This can be seen in the 
top right panel of Fig.~\ref{fig:rhor}, in which we plot the values and uncertainties in $V_{\rm max}$ for each dSph, as inferred from the MCMC 
chains used in Ref.~\cite{Pace:2018tin}.
Even with the prior imposed, the dSphs have essentially consistent $V_{\rm max}$ to within the uncertainties.

These two facts together explain why 
the maximized likelihood tends to favor the model 
of Sommerfeld-enhanced annihilation (given current 
$J$-factor uncertainties) even if the true model 
is $s$-wave annihilation.  Consider two annihilation models, 1 and 2, with velocity-dependence specified by $n_1$ and $n_2$ such that $n_1-n_2=\Delta n$.  The $J$-factor of a dSph computed assuming model 2 will differ from that assuming model 1 by a factor of order $V_{\rm max}^{\Delta n}$. Since the $V_{\rm max}$ values are similar across all dSphs, if $\Delta n$ is small, then the $J$-factors of all dSphs for
these two models will roughly differ by only an overall common factor, which can 
be compensated by a rescaling of $\Phi_{PP}$.  Moreover, if $n_1 > n_2$, then as we argued above, the $n_1$ model will yield larger uncertainties on the $J$-factors.  The large $J$-factor uncertainties mean that when varying $\Phi_{PP}$, model 1 (with the larger value of $n$) will yield a lower maximum likelihood than model 2, regardless of which was the true model.  Therefore, the larger $n$ model ($n_1$) will be disfavored.  For large $\Delta n$, on the other hand, the small differences in $V_{\rm max}$ between the different dSphs will be magnified.  If $\Delta n$ is sufficiently large, the $J$-factors of the different dSphs will become sufficiently different that the true $n$ model will be favored despite possible differences in the $J$-factor uncertainties.

This effect explains the strange behavior seen in Fig.~\ref{fig:deltaloglike_CTA_sdefault} when the data generated assuming $s$-wave annihilation are analyzed with the Sommerfeld model.  In this case, $\Delta n = 1$ and the model with the lower value of $n$ (Sommerfeld) is preferred slightly over $s$-wave for current $J$-factor uncertainty, even though this is not the true model. This effect persists even at high $\Phi_{PP}$, since in these cases the $J$-factor uncertainty dominates the width of the likelihood.  The only remedy to this situation is to obtain tighter constraints on the $J$-factors, as seen in the bottom row of Fig.~\ref{fig:deltaloglike_CTA_sdefault}.  
Note that, as shown in Fig.~\ref{fig:deltaloglike_CTA_somdefault}, the Sommerfeld model is always preferred when the data are generated assuming Sommerfeld annihilation.  Since the Sommerfeld model has the lowest value of $n$, other annihilation models will yield larger $J$-factor uncertainties; thus, in this case, the true model will also have the smallest $J$-factor 
uncertainties.

Similarly, we see in Figure~\ref{fig:deltaloglike_CTA_pdefault} that, if 
the mock data are generated assuming $p$-wave annihilation, 
then with current $J$-factor uncertainties, the likelihood 
would show a preference for the $s$-wave model over the 
true $p$-wave model.  But the true model is preferred over 
the Sommerfeld model; although the $J$-factors for the 
Sommerfeld model are smaller, for this case $\Delta n$ is 
large enough that the relative differences in the $J$-factors 
can be distinguished.  But in all cases we find that, if the 
$J$-factor uncertainties can be sufficiently reduced, then 
a reasonable data set can be used to distinguish the 
true model of dark matter annihilation.

One might expect imposing the prior on the $r_s$-$\rho_s$ relation to help here, since this prior will decrease the $J$-factor uncertainty (as seen for most dSphs in the bottom right panel of Fig.~\ref{fig:rhor}).  However, we find that the $V_{\rm max}$ values still remain close together (to within the uncertainties) upon the application of the $r_s$-$\rho_s$ prior, as seen in the top right panel of Fig.~\ref{fig:rhor}.  Furthermore, we find that imposing the cosmological prior can shift the mean $J$-factors.  As seen in Appendix~\ref{app:J_constraints}, the imposition of the prior tends to reduce the $J$-factors more as $n$ is increased.  This explains why when the data are generated assuming $s$-wave and $p$-wave annihilation (Figs.~\ref{fig:deltaloglike_CTA_sdefault} and \ref{fig:deltaloglike_CTA_pdefault}, respectively), the imposition of the prior typically leads to slightly reduced $\Delta \ln \mathcal{L}_{\rm max}$, while for Sommerfeld-enhanced annihilation (Fig.~\ref{fig:deltaloglike_CTA_somdefault}), the imposition of the prior leads to somewhat enhanced $\Delta \ln \mathcal{L}_{\rm max}$. 

\section{Conclusions}
\label{sec:discussion}

We have considered the prospects for gamma-ray searches of dwarf spheroidal 
galaxies to determine the power-law 
velocity-dependence of the dark matter annihilation 
cross section.  The key principle behind this study is that if the dark 
matter profile is parameterized only by a scale radius $r_s$ and scale 
density $\rho_s$, then the dark matter velocity distribution in any subhalo 
is characterized by a single velocity parameter $V_{\rm max} \propto 
(G_N \rho_s r_s^2)^{1/2}$.  Thus, 
the photon flux from any dSph is proportional to powers of $(\rho_s r_s^2)^{1/2}$, 
where the proportionality constant is universal, but $\rho_s$ and $r_s$ are 
unique to each dSph and can be estimated from stellar data.  Although the 
photon flux from one dSph cannot distinguish the effect of the velocity-dependence 
from that of the overall normalization of the annihilation flux, the relative photon fluxes from many 
dSphs should, in principle, be sufficient to distinguish between different 
models of dark matter annihilation.  

In practice, we have found that this intuition is correct, but with some caveats.  We have considered theoretically-motivated scenarios in which the annihilation 
cross section has a velocity-dependence proportional to $(v/c)^n$, with 
$n=-1, 0, 2, 4$.  In general, more exposure is required to reject a dark matter 
annihilation model with the wrong velocity-dependence than is required to reject 
the background-only scenario.  But the larger the difference in $n$ between the 
true model and the alternate hypothesis, the smaller the exposure required to reject the 
false hypothesis.

Interestingly, for current $J$-factor uncertainty levels, we have found that if the true velocity dependence of the annihilation is $n_{\rm true}$, it can be difficult or impossible to reject models with $n < n_{\rm true}$, even at large exposure and $\Phi_{PP}$.  The basic reason is 
that uncertainties in the velocity-dependent $J$-factor tend to increase with 
$n$, given the stellar velocity data.  Because the velocity parameters of the various dSphs which are currently observed are all roughly ${\cal O}(10~\km/\s)$, up to uncertainties, the 
$V_{\rm max}$-dependent rescaling of the $J$-factors which would be required for a 
different choice of annihilation model is approximately the same for all dSphs, 
when compared to their current uncertainties.  This rescaling can be absorbed by the 
overall normalization $\Phi_{PP}$.  Thus, if the likelihood is dominated 
by the uncertainties in the $J$-factors, the large $J$-factor uncertainties for the large $n$ models can cause these models to be disfavored in a likelihood analysis.

But we also see that, if the future surveys can reduce the uncertainty in the 
$J$-factor, then one could realistically distinguish the velocity-dependence of 
the dark matter annihilation cross section, with an exposure only modestly larger
than that needed to reject the background-only model.  We have shown that the necessary reduction in $J$-factor uncertainties can be achieved with future stellar velocity measurements that probe fainter magnitude stars.  However, we also see that current levels of systematic error in the $J$-factor determination will make determining the velocity-dependence of the annihilation significantly more difficult in several cases.  Our analysis motivates additional efforts to reduce these systematic errors.  
For example, the systemic error from the unknown stellar anisotropy can be probed with future tangential velocity measurements \citep[e.g.,][]{Massari2018NatAs...2..156M},  the triaxiality of DM halos could be addressed from numerical simulations \citep{Munoz:2011}, and the DM velocity anisotropy can potentially also be addressed from numerical simulations \citep{Board2021JCAP...04..070B}.  

In summary, upcoming gamma-ray observations of dSphs may not only be able to detect the presence of dark matter annihilation, but may also be able to determine  the velocity-dependence of the annihilation cross section.  But an improvement in the precision of stellar data and control of systematic errors in the $J$-factor determination would be required in order for the latter determination to be robust.

\vspace{0.5cm}

{\bf Acknowledgements}

We are grateful to Danny Marfatia, Emmanuel Moulin and 
Tracy Slatyer for useful discussions.
The work of JK is supported in part by DOE grant DE-SC0010504.
ABP is supported by NSF grant AST-1813881.
JR is supported by NSF grant AST-1934744.

\bibliographystyle{JHEP}
\bibliography{thebib}

\appendix

\section{$J$-factors with cosmological prior}
\label{app:J_constraints}

\begin{figure}
    \centering
    \includegraphics[scale=0.7]{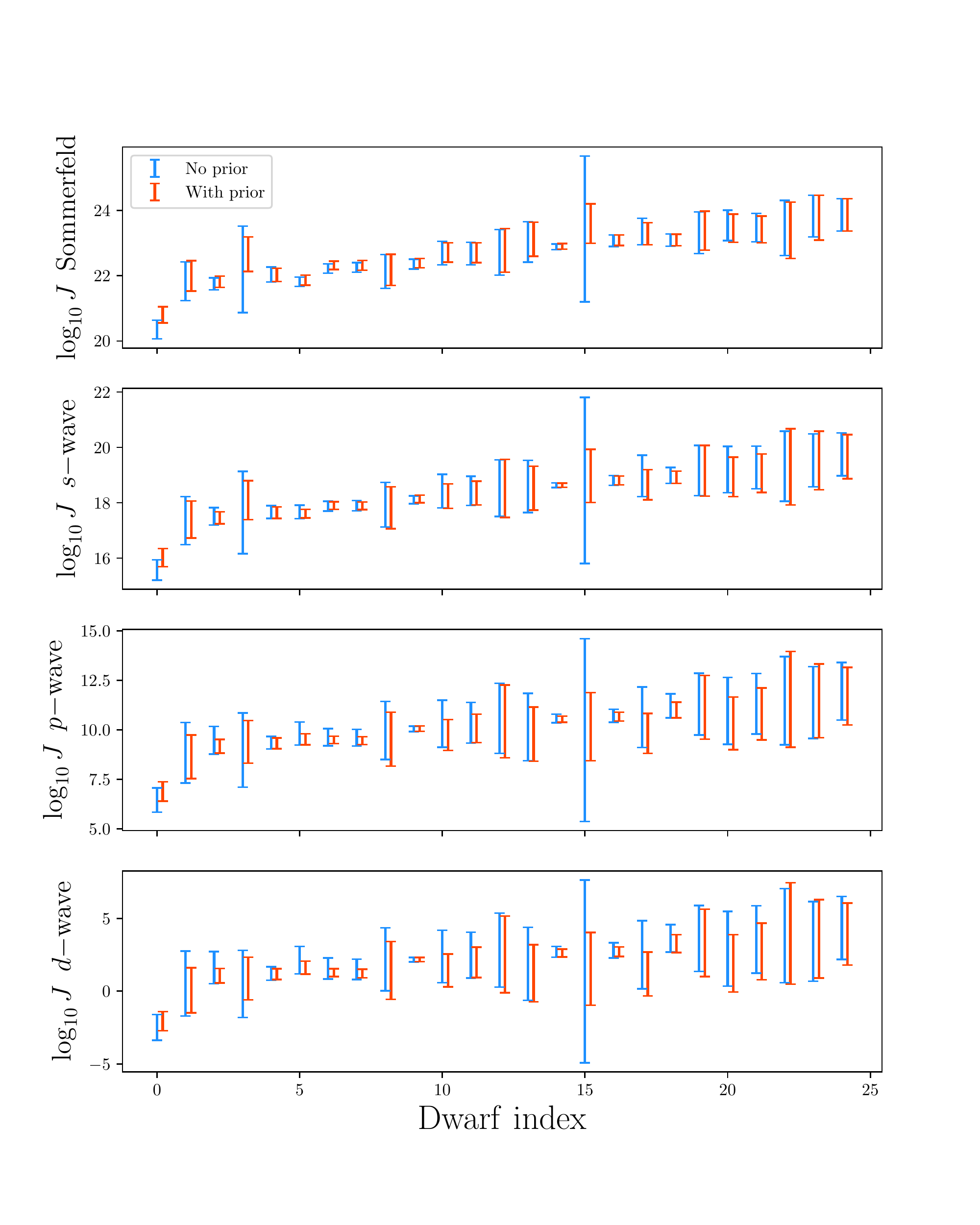}
    \caption{$J$-factors computed assuming different velocity-dependence for the dark matter annihilation cross section, with and without the imposition of the cosmological prior discussed in \S\ref{sec:Jwithprior}.  $J$ is in units of ${\rm GeV}^2 {\rm cm}^{-5}$, and the dSphs are ordered by their $s$-wave $J$-factors.}
    \label{fig:J_allanntype}
\end{figure}

In Table~\ref{tab:Jfactors}, we present the posteriors on the dSph $J$-factors derived from the analysis of \cite{Boddy:2020}, with and without the cosmological prior discussed in \S\ref{sec:Jwithprior}.  We plot these results in Fig.~\ref{fig:J_allanntype}.

\begin{table}
\tiny
\renewcommand\arraystretch{1.7}
\begin{tabular}{|l|p{0.07\textwidth}|p{0.07\textwidth}|p{0.07\textwidth}|p{0.07\textwidth}|p{0.09\textwidth}|p{0.09\textwidth}|p{0.09\textwidth}|p{0.09\textwidth}|}
\hline
dSph name & Som. no prior & Som. w/ prior & $s$ no prior & $s$ w/ prior & $p$ no prior & $p$ w/ prior & $d$ no prior & $d$ w/ prior \\ \hline
\hline
aquarius2 & $22.70^{ +0.44 }_{ -0.48 } $ & $22.73^{ +0.37 }_{ -0.45 } $ & $18.48^{ +0.62 }_{ - 0.72 } $ & $18.42^{ + 0.56 }_{ -0.78 } $ & $10.51^{ +1.08 }_{ -1.27 } $ & $10.23^{ +0.93 }_{ -1.56 } $ & $2.70^{ +1.56 }_{ -1.86 } $ & $2.22^{ + 1.35 }_{ -2.35 } $ \\
\hline
bootes1 & $22.68^{ +0.23 }_{ -0.24 } $ & $22.71^{ +0.19 }_{ -0.22 } $ & $18.39^{ +0.38 }_{ - 0.45 } $ & $18.22^{ + 0.27 }_{ -0.61 } $ & $10.24^{ +0.76 }_{ -0.93 } $ & $9.69^{ +0.43 }_{ -1.49 } $ & $2.29^{ +1.17 }_{ -1.42 } $ & $1.33^{ + 0.60 }_{ -2.37 } $ \\
\hline
canesvenatici1 & $21.75^{ +0.12 }_{ -0.12 } $ & $21.82^{ +0.11 }_{ -0.05 } $ & $17.49^{ +0.17 }_{ - 0.24 } $ & $17.45^{ + 0.13 }_{ -0.27 } $ & $9.38^{ +0.34 }_{ -0.63 } $ & $9.16^{ +0.21 }_{ -0.85 } $ & $1.45^{ +0.54 }_{ -1.04 } $ & $1.03^{ + 0.28 }_{ -1.45 } $ \\
\hline
canesvenatici2 & $22.12^{ +0.33 }_{ -0.35 } $ & $22.17^{ +0.30 }_{ -0.30 } $ & $17.92^{ +0.53 }_{ - 0.55 } $ & $17.78^{ + 0.44 }_{ -0.69 } $ & $9.95^{ +0.97 }_{ -1.03 } $ & $9.43^{ +0.76 }_{ -1.55 } $ & $2.16^{ +1.46 }_{ -1.52 } $ & $1.27^{ + 1.09 }_{ -2.42 } $ \\
\hline
carina2 & $23.03^{ +0.39 }_{ -0.41 } $ & $23.10^{ +0.32 }_{ -0.33 } $ & $18.57^{ +0.59 }_{ - 0.66 } $ & $18.49^{ + 0.46 }_{ -0.73 } $ & $10.06^{ +1.05 }_{ -1.29 } $ & $9.70^{ +0.77 }_{ -1.66 } $ & $1.73^{ +1.51 }_{ -1.93 } $ & $1.08^{ + 1.07 }_{ -2.57 } $ \\
\hline
carina & $22.25^{ +0.09 }_{ -0.11 } $ & $22.32^{ +0.10 }_{ -0.03 } $ & $17.88^{ +0.10 }_{ - 0.11 } $ & $17.89^{ + 0.09 }_{ -0.11 } $ & $9.54^{ +0.16 }_{ -0.29 } $ & $9.45^{ +0.11 }_{ -0.38 } $ & $1.37^{ +0.26 }_{ -0.50 } $ & $1.19^{ + 0.15 }_{ -0.69 } $ \\
\hline
comaberenices & $23.46^{ +0.28 }_{ -0.29 } $ & $23.40^{ +0.25 }_{ -0.36 } $ & $19.25^{ +0.50 }_{ - 0.57 } $ & $19.01^{ + 0.38 }_{ -0.81 } $ & $11.29^{ +1.05 }_{ -1.15 } $ & $10.69^{ +0.70 }_{ -1.75 } $ & $3.50^{ +1.62 }_{ -1.75 } $ & $2.54^{ + 1.03 }_{ -2.71 } $ \\
\hline
crater2 & $20.34^{ +0.18 }_{ -0.20 } $ & $20.81^{ +0.17 }_{ -0.27 } $ & $15.56^{ +0.23 }_{ - 0.25 } $ & $16.03^{ + 0.24 }_{ -0.23 } $ & $6.43^{ +0.37 }_{ -0.42 } $ & $6.91^{ +0.35 }_{ -0.06 } $ & $-2.54^{ +0.54 }_{ -0.63 } $ & $-2.05^{ + 0.47 }_{ -0.14 } $ \\
\hline
draco1 & $23.08^{ +0.12 }_{ -0.13 } $ & $23.09^{ +0.11 }_{ -0.12 } $ & $18.96^{ +0.16 }_{ - 0.20 } $ & $18.91^{ + 0.13 }_{ -0.25 } $ & $11.13^{ +0.30 }_{ -0.47 } $ & $10.97^{ +0.20 }_{ -0.64 } $ & $3.50^{ +0.46 }_{ -0.75 } $ & $3.20^{ + 0.31 }_{ -1.05 } $ \\
\hline
fornax & $22.36^{ +0.10 }_{ -0.10 } $ & $22.39^{ +0.09 }_{ -0.07 } $ & $18.11^{ +0.09 }_{ - 0.10 } $ & $18.13^{ + 0.09 }_{ -0.07 } $ & $10.05^{ +0.09 }_{ -0.09 } $ & $10.06^{ +0.09 }_{ -0.08 } $ & $2.16^{ +0.10 }_{ -0.10 } $ & $2.17^{ + 0.09 }_{ -0.10 } $ \\
\hline
hercules & $21.84^{ +0.39 }_{ -0.38 } $ & $21.98^{ +0.29 }_{ -0.23 } $ & $17.35^{ +0.54 }_{ - 0.58 } $ & $17.38^{ + 0.41 }_{ -0.55 } $ & $8.78^{ +0.91 }_{ -1.10 } $ & $8.59^{ +0.66 }_{ -1.29 } $ & $0.38^{ +1.31 }_{ -1.68 } $ & $0.01^{ + 0.94 }_{ -2.05 } $ \\
\hline
horologium1 & $23.42^{ +0.51 }_{ -0.59 } $ & $23.36^{ +0.55 }_{ -0.65 } $ & $19.28^{ +0.80 }_{ - 0.87 } $ & $19.25^{ + 0.90 }_{ -0.89 } $ & $11.39^{ +1.39 }_{ -1.59 } $ & $11.47^{ +1.59 }_{ -1.51 } $ & $3.67^{ +2.03 }_{ -2.39 } $ & $3.85^{ + 2.31 }_{ -2.20 } $ \\
\hline
hydrus1 & $23.34^{ +0.25 }_{ -0.28 } $ & $23.27^{ +0.20 }_{ -0.35 } $ & $18.93^{ +0.47 }_{ - 0.57 } $ & $18.61^{ + 0.29 }_{ -0.89 } $ & $10.58^{ +1.03 }_{ -1.21 } $ & $9.73^{ +0.51 }_{ -2.07 } $ & $2.42^{ +1.61 }_{ -1.85 } $ & $1.01^{ + 0.75 }_{ -3.25 } $ \\
\hline
leo1 & $21.81^{ +0.09 }_{ -0.10 } $ & $21.86^{ +0.09 }_{ -0.05 } $ & $17.64^{ +0.12 }_{ - 0.19 } $ & $17.60^{ + 0.09 }_{ -0.23 } $ & $9.72^{ +0.27 }_{ -0.53 } $ & $9.50^{ +0.15 }_{ -0.75 } $ & $2.00^{ +0.47 }_{ -0.85 } $ & $1.57^{ + 0.22 }_{ -1.28 } $ \\
\hline
leo2 & $22.03^{ +0.15 }_{ -0.16 } $ & $22.02^{ +0.13 }_{ -0.17 } $ & $17.66^{ +0.15 }_{ - 0.16 } $ & $17.64^{ + 0.14 }_{ -0.17 } $ & $9.33^{ +0.18 }_{ -0.20 } $ & $9.31^{ +0.18 }_{ -0.22 } $ & $1.19^{ +0.23 }_{ -0.26 } $ & $1.16^{ + 0.23 }_{ -0.29 } $ \\
\hline
reticulum2 & $23.53^{ +0.30 }_{ -0.32 } $ & $23.43^{ +0.25 }_{ -0.42 } $ & $19.16^{ +0.53 }_{ - 0.64 } $ & $18.87^{ + 0.37 }_{ -0.93 } $ & $10.86^{ +1.08 }_{ -1.38 } $ & $10.19^{ +0.67 }_{ -2.05 } $ & $2.75^{ +1.66 }_{ -2.13 } $ & $1.68^{ + 0.99 }_{ -3.19 } $ \\
\hline
sagittarius2 & $22.03^{ +0.70 }_{ -1.16 } $ & $22.66^{ +0.34 }_{ -0.53 } $ & $17.48^{ +0.79 }_{ - 1.23 } $ & $18.09^{ + 0.46 }_{ -0.63 } $ & $8.83^{ +1.07 }_{ -1.42 } $ & $9.37^{ +0.69 }_{ -0.88 } $ & $0.35^{ +1.37 }_{ -1.66 } $ & $0.83^{ + 0.93 }_{ -1.19 } $ \\
\hline
sculptor & $22.88^{ +0.05 }_{ -0.06 } $ & $22.89^{ +0.05 }_{ -0.04 } $ & $18.63^{ +0.05 }_{ - 0.05 } $ & $18.63^{ + 0.05 }_{ -0.06 } $ & $10.55^{ +0.09 }_{ -0.15 } $ & $10.52^{ +0.08 }_{ -0.18 } $ & $2.65^{ +0.16 }_{ -0.26 } $ & $2.59^{ + 0.13 }_{ -0.32 } $ \\
\hline
segue1 & $23.71^{ +0.53 }_{ -0.39 } $ & $23.60^{ +0.37 }_{ -0.49 } $ & $19.12^{ +0.68 }_{ - 0.63 } $ & $18.96^{ + 0.58 }_{ -0.79 } $ & $10.32^{ +1.08 }_{ -1.40 } $ & $10.11^{ +0.98 }_{ -1.60 } $ & $1.67^{ +1.50 }_{ -2.30 } $ & $1.44^{ + 1.42 }_{ -2.53 } $ \\
\hline
sextans & $22.21^{ +0.09 }_{ -0.10 } $ & $22.32^{ +0.09 }_{ -0.00 } $ & $17.87^{ +0.10 }_{ - 0.12 } $ & $17.90^{ + 0.09 }_{ -0.09 } $ & $9.56^{ +0.17 }_{ -0.33 } $ & $9.49^{ +0.12 }_{ -0.40 } $ & $1.42^{ +0.26 }_{ -0.61 } $ & $1.26^{ + 0.16 }_{ -0.77 } $ \\
\hline
tucana2 & $23.30^{ +0.39 }_{ -0.44 } $ & $23.35^{ +0.35 }_{ -0.39 } $ & $19.13^{ +0.56 }_{ - 0.65 } $ & $19.09^{ + 0.52 }_{ -0.70 } $ & $11.24^{ +0.99 }_{ -1.13 } $ & $11.00^{ +0.87 }_{ -1.37 } $ & $3.53^{ +1.44 }_{ -1.66 } $ & $3.08^{ + 1.23 }_{ -2.11 } $ \\
\hline
ursamajor1 & $22.68^{ +0.23 }_{ -0.22 } $ & $22.70^{ +0.20 }_{ -0.20 } $ & $18.40^{ +0.32 }_{ - 0.37 } $ & $18.33^{ + 0.26 }_{ -0.44 } $ & $10.26^{ +0.55 }_{ -0.82 } $ & $10.03^{ +0.40 }_{ -1.05 } $ & $2.27^{ +0.76 }_{ -1.36 } $ & $1.91^{ + 0.57 }_{ -1.72 } $ \\
\hline
ursamajor2 & $23.85^{ +0.32 }_{ -0.33 } $ & $23.84^{ +0.30 }_{ -0.34 } $ & $19.72^{ +0.49 }_{ - 0.54 } $ & $19.62^{ + 0.46 }_{ -0.65 } $ & $11.90^{ +0.93 }_{ -1.08 } $ & $11.58^{ +0.80 }_{ -1.39 } $ & $4.25^{ +1.39 }_{ -1.64 } $ & $3.72^{ + 1.16 }_{ -2.17 } $ \\
\hline
ursaminor & $23.07^{ +0.12 }_{ -0.12 } $ & $23.09^{ +0.11 }_{ -0.10 } $ & $18.80^{ +0.11 }_{ - 0.11 } $ & $18.80^{ + 0.10 }_{ -0.11 } $ & $10.68^{ +0.14 }_{ -0.18 } $ & $10.66^{ +0.13 }_{ -0.20 } $ & $2.73^{ +0.19 }_{ -0.29 } $ & $2.69^{ + 0.18 }_{ -0.33 } $ \\
\hline
willman1 & $23.82^{ +0.39 }_{ -0.42 } $ & $23.74^{ +0.41 }_{ -0.49 } $ & $19.46^{ +0.52 }_{ - 0.73 } $ & $19.47^{ + 0.62 }_{ -0.72 } $ & $11.14^{ +0.89 }_{ -1.60 } $ & $11.36^{ +1.09 }_{ -1.38 } $ & $3.01^{ +1.32 }_{ -2.47 } $ & $3.42^{ + 1.59 }_{ -2.06 } $ \\
\hline
\end{tabular}
\caption{$J$-factors computed from the analysis of stellar data in \cite{Boddy:2020} with and without imposing a cosmological prior on the $r_s$-$\rho_s$ relation (see \S\ref{sec:Jwithprior}).  Numbers represent $\log_{10}(J/{\rm GeV}^2 {\rm cm}^{-5})$.}
\label{tab:Jfactors}
\end{table}
\clearpage

\clearpage

\section{Likelihood results for different observational configurations and velocity dependences}
\label{app:alternate_results}

\begin{figure}
    \centering
    \includegraphics[scale=0.55]{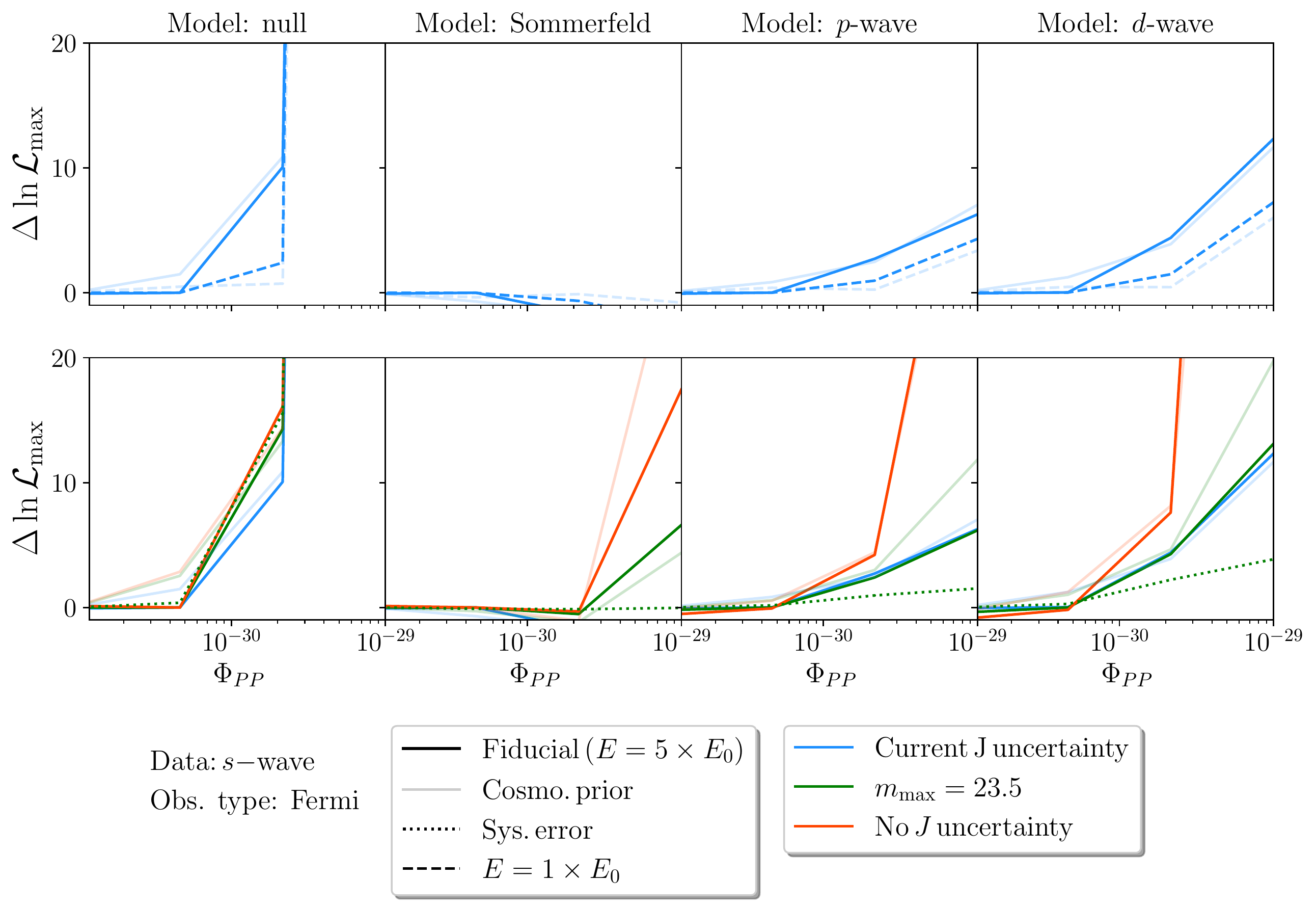}
    \caption{Likelihood results when mock data are generated assuming $s$-wave annihilation and Fermi-like observations.  The units of $\Phi_{PP}$ are $\cm^3{\rm s}^{-1} {\rm GeV}^{-2} $.}
    \label{fig:deltaloglike_fermi_sdefault}
\end{figure}

\begin{figure}
    \centering
    \includegraphics[scale=0.55]{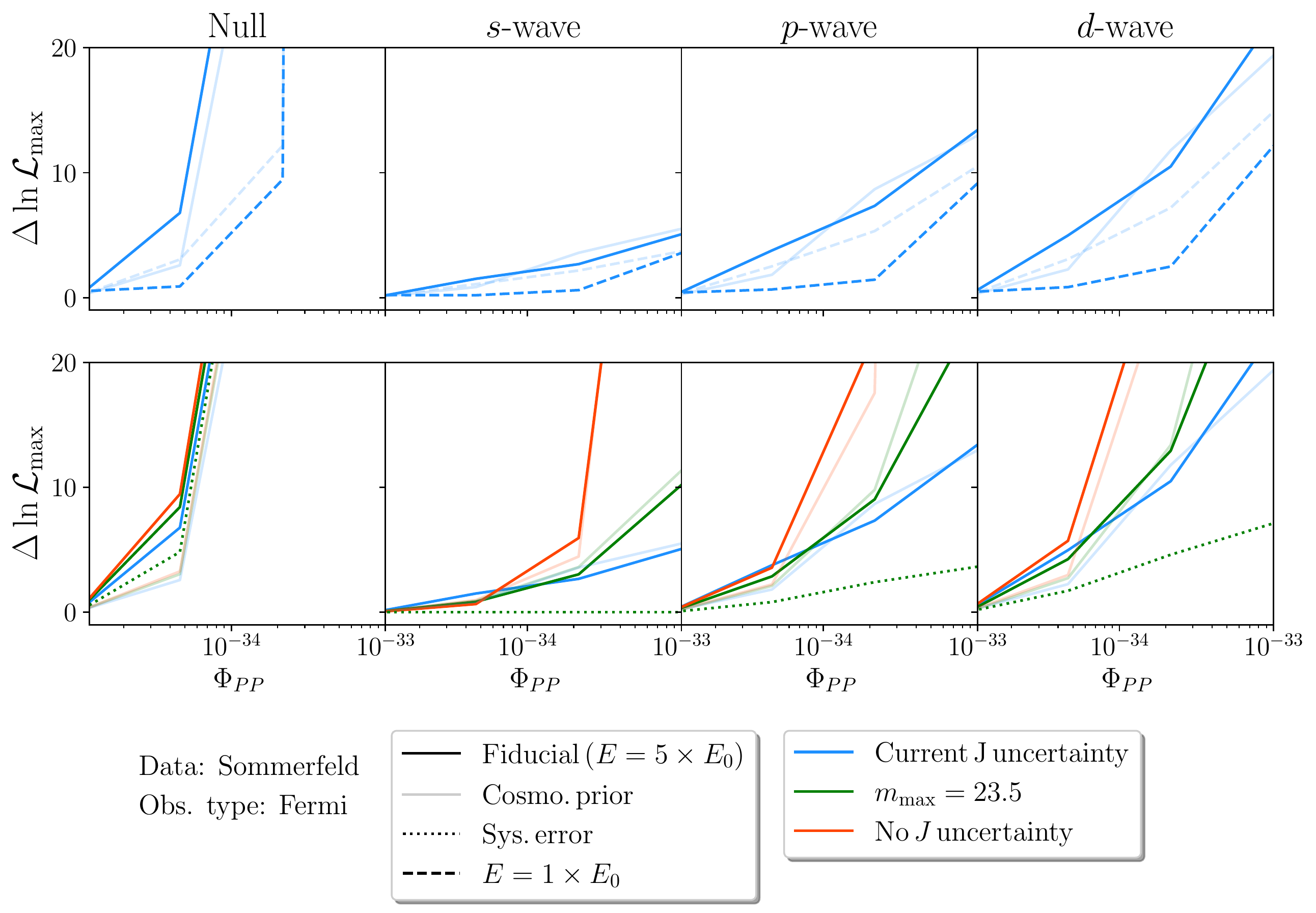}
    \caption{Likelihood results when mock data are generated assuming Sommerfeld-enhanced annihilation and Fermi-like observations.  The units of $\Phi_{PP}$ are $\cm^3{\rm s}^{-1} {\rm GeV}^{-2}$.}
    \label{fig:deltaloglike_fermi_somdefault}
\end{figure}

\begin{figure}
    \centering
    \includegraphics[scale=0.55]{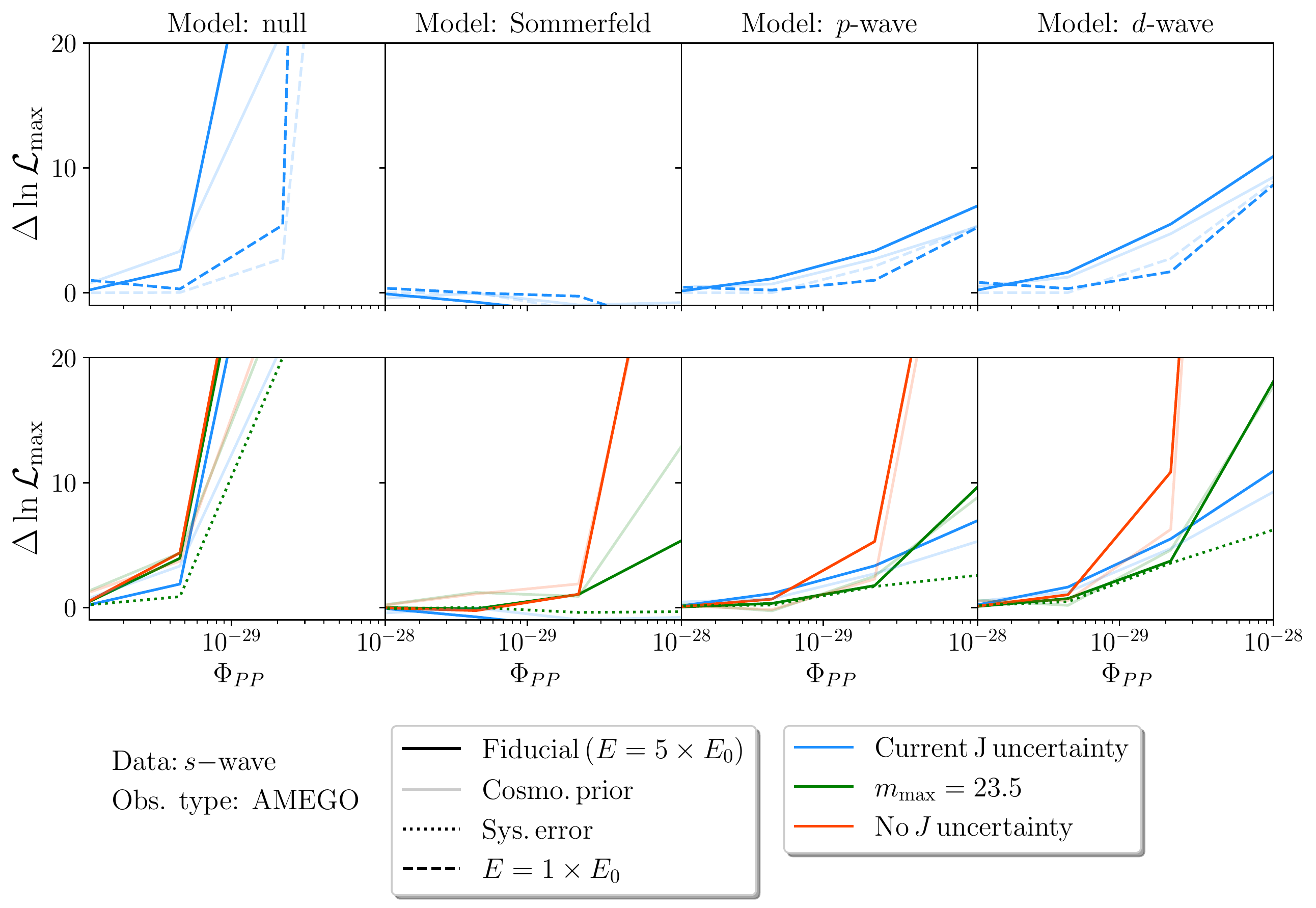}
    \caption{Likelihood results when mock data are generated assuming $s$-wave annihilation and AMEGO-like observations.  The units of $\Phi_{PP}$ are $\cm^3
    {\rm s}^{-1} {\rm GeV}^{-2} $.}
    \label{fig:deltaloglike_AMEGO_sdefault}
\end{figure}

\begin{figure}
    \centering
    \includegraphics[scale=0.55]{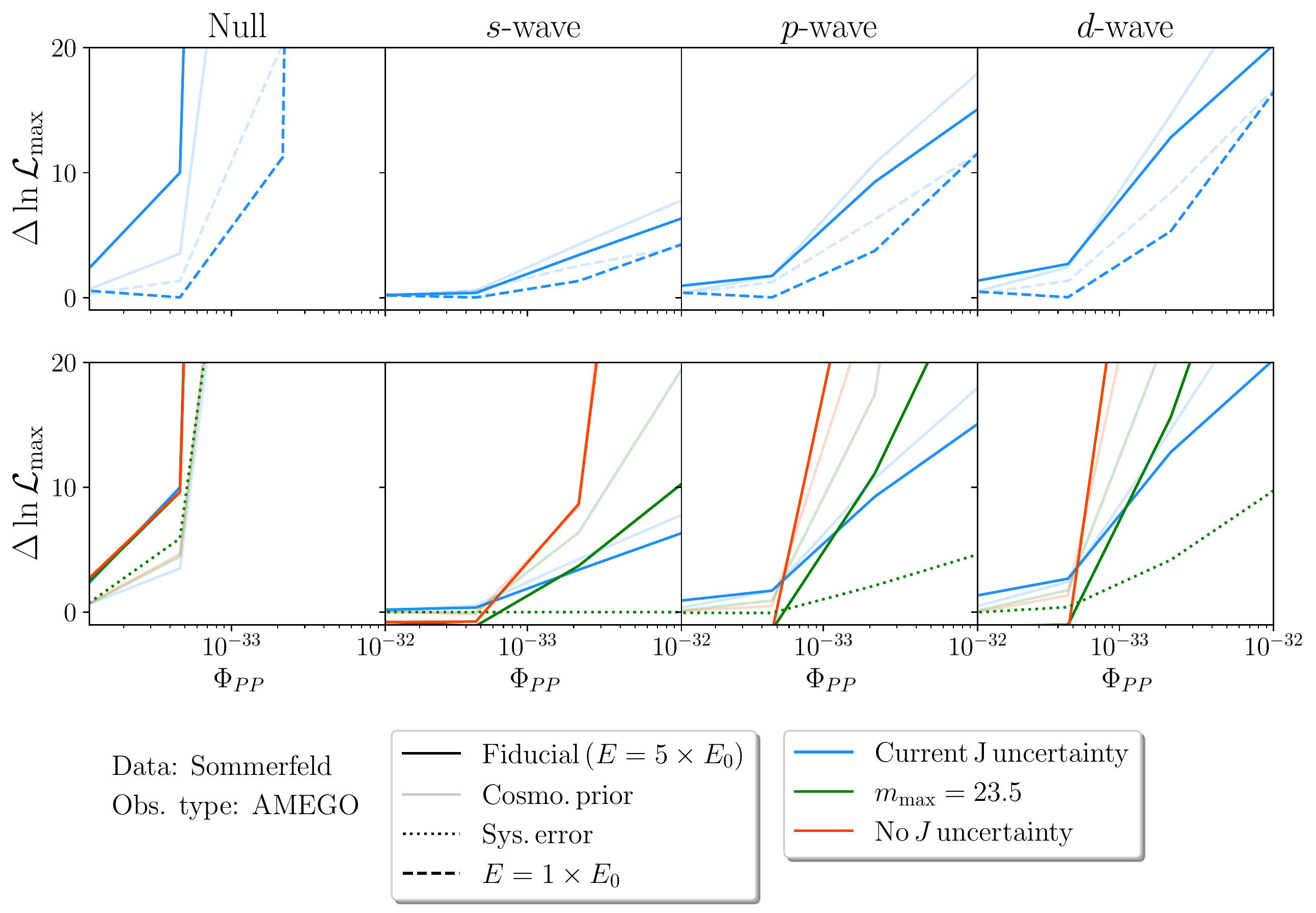}
    \caption{Likelihood results when mock data are generated assuming Sommerfeld-enhanced annihilation and AMEGO-like observations.  The units of $\Phi_{PP}$ are $\cm^3 {\rm s}^{-1} {\rm GeV}^{-2} $.}
    \label{fig:deltaloglike_AMEGO_somdefault}
\end{figure}

In Figs.~\ref{fig:deltaloglike_fermi_sdefault} through \ref{fig:deltaloglike_AMEGO_somdefault} we present the results of our likelihood analysis for different experimental configurations and for data generated assuming different models for the dark matter annihilation velocity dependence.

In Figure~\ref{fig:deltaloglike_fermi_sdefault} 
(Figure~\ref{fig:deltaloglike_fermi_somdefault}), we 
assume that the true model is $s$-wave (Sommerfeld-enhanced) annihilation, and assume a Fermi-like experimental configuration, 
as discussed in Subsection~\ref{sec:background_modeling}.  
Note that, for an exposure similar to the current 
Fermi exposure, the $s$-wave (Sommerfeld) model can 
be distinguished 
from the background model if 
$\Phi_{PP} \geq {\cal O}(10^{-30}) \cm^3 \s^{-1} \gev^{-2}$ 
($\Phi_{PP} \geq {\cal O}(10^{-34}) \cm^3 \s^{-1} 
\gev^{-2}$), 
consistent with the results found in Ref.~\cite{Boddy:2018qur,Boddy:2019kuw}.
Similarly, in  Figure~\ref{fig:deltaloglike_AMEGO_sdefault} 
(Figure~\ref{fig:deltaloglike_AMEGO_somdefault}), we 
assume that the true model is $s$-wave (Sommerfeld-enhanced) annihilation, and assume a AMEGO-like experimental configuration.

\end{document}